%% file: bpmn_paper_verification.tex
\documentclass[preprint,12pt]{elsarticle}

\usepackage{graphicx}%
\usepackage{multirow}%
\usepackage{amsmath,amssymb,amsfonts}%
\usepackage{mathrsfs}%
\usepackage{algorithm}%
\usepackage{algorithmicx}%
\usepackage{algpseudocode}%
\usepackage{algpseudocode}%
\usepackage{url}%

\newtheorem{theorem}{Theorem}[section]

\usepackage{verbatim}
\usepackage{acronym}
\acrodef{BPMN}[BPMN]{Business Process Model and Notation}
\acrodef{DMN}[DMN]{Decision Model and Notation}
\acrodef{FEEL}[FEEL]{Friendly Enough Expression Language}
\acrodef{OMG}[OMG]{Object Management Group}
\acrodef{MCDC}[MC/DC]{Modified Condition/Decision Criterion}
\acrodef{IT}[IT]{Information Technology}
\acrodef{DENEB}[DENEB]{Development and Execution of iNteroperable dynamic wEB processes}
\acrodef{IoT}[IoT]{Internet of Things}
\acrodef{SysML}[SysML]{Systems Modeling Language}

\newcommand{\BDTrans}{\textsf{BDTrans}\ }
\newcommand{\BDTest}{\textsf{BDTest}\ }
\newcommand{\BDTransTest}{\textsf{BDTransTest}\ }
\newcommand{\FEELJava}{\textsf{FEEL2Java}\ }
\newcommand{\BeL}{\textsf{BeL}\ }
\newcommand{\BDTransNoSp}{\textsf{BDTrans}}
\newcommand{\BDTestNoSp}{\textsf{BDTest}}
\newcommand{\BDTransTestNoSp}{\textsf{BDTransTest}}

\newcommand{\BeLNoSp}{\textsf{BeL}}
\newcommand{\FunctionDef}[1]{\textsl{#1}}
\newcommand{\Example}[1]{\textsl{#1}}
\newcommand{\github}{\url{https://github.com/gdellapenna/BPMNModelTranslator}}

\usepackage{listings}
\journal{Journal of Systems \& Software}

\begin{document}

\begin{frontmatter}

  \title{Automating Execution and Verification of BPMN+DMN Business Processes}

\author[univaq]{Giuseppe Della Penna}
\ead{giuseppe.dellapenna@univaq.it}
\author[univaq]{Igor Melatti}
\ead{igor.melatti@univaq.it}

\affiliation[univaq]{organization={Department of Information Engineering, Computer Science and Mathematics, University of L'Aquila},
  addressline={Via Vetoio 1},
  city={L'Aquila},
  postcode={67100},
  country={Italy}}



\begin{keyword}



  BPMN processes \sep DMN tables \sep Business Processes Verification
\end{keyword}

\begin{abstract}
  The increasing and widespread use of BPMN business processes, also embodying DMN tables, requires tools and methodologies to verify their correctness. However, most commonly used frameworks to build BPMN+DMN models only allow designers to detect syntactical errors, thus ignoring semantic (behavioural) faults. This forces business processes designers to manually run single executions of their BPMN+DMN processes using proprietary tools in order to detect failures. Furthermore, how proprietary tools translate a BPMN+DMN process to a computer simulation is left unspecified.
  In this paper, we advance this state of the art by designing a tool, named \textsf{BDTransTest} providing: i) a translation from a BPMN + DMN process ${\cal B}$ to a Java program $P$; 
  ii) the synthesis and execution of a testing plan for ${\cal B}$, that may require the business designer to disambiguate some input domain; iii) the analysis of the coverage achieved by the testing plan in terms of nodes and edges of  ${\cal B}$.
  Finally, we provide an experimental evaluation of our methodology on BPMN+DMN processes from the literature.
\end{abstract}
\end{frontmatter}

\input{intro.tex}

\input{related_work.tex}

\input{background.tex}

\input{translation.tex}
\input{post_processing.tex}

\input{execution.tex}

\input{conclusion.tex}

 \bibliographystyle{elsarticle-num} 
 \bibliography{bpmn.bib}

\end{document}

%% file: intro.tex
\section{Introduction}\label{sec:introduction}

Since its first release dating back to 2004, \ac{BPMN}~\citep{bpmn,bpmn_book}
has gained an ever-increasing importance in modeling business processes in companies all around the world~\citep{CCFR23,OFRMM12}.
As an example, BPMN.iO reports nearly 8000 currently active \ac{BPMN} projects per week, and there exist tens of publicly available repositories maintaining thousands of \ac{BPMN}s each.

Starting from 2015, the \ac{DMN}~\citep{dmn,DFM16}
has been designed to complement \ac{BPMN} on decision making, by providing a graphical notation and an expression language to describe in a non-specialistic way possibly complex Boolean formulas. Also in this case, usage of the \ac{BPMN}+\ac{DMN} combo has been high since the first beginning, and is currently growing~\cite{HSS20,HEB21}.

Unfortunately, as is common with any type of ``programming'', it is rather easy to make errors when describing a business process using \ac{BPMN} and DNM notations. However, the most commonly used frameworks to build \ac{BPMN}+\ac{DMN} models (e.g., Camunda \footnote{\url{https://camunda.com/}}) only allow the designer to detect syntactical errors. This excludes the various types of semantic faults that a \ac{BPMN}+\ac{DMN} model can be tampered with: unexpected behaviors, nontermination, unreachable \ac{BPMN} blocks, and so on. Furthermore, even the simple ``execution'' of a given \ac{BPMN}+\ac{DMN} could be not easy, as it is typically required to use some proprietary framework (e.g., again Camunda) which only allows one execution at a time and does not provide details about the inner translation from \ac{BPMN}+\ac{DMN} to a software process.

To the best of our knowledge, there are no works in the literature that tackle the problem of automatically translating a given business process, described using \ac{BPMN} and \ac{DMN} notations, into a software process in such a way that the software code can be inspected by a software engineer. Furthermore, while many works in the literature address the problem of verifying a \ac{BPMN} business process, the verification of a given business process, described using both \ac{BPMN} and \ac{DMN} notations, has attracted less attention (see Section~\ref{related_work.sec}). 

In this paper, we show how both obstructions outlined above may be tackled together. Namely, we achieve the following contributions.

\begin{figure}
\centerline{\framebox{\includegraphics[height=0.6\textheight]{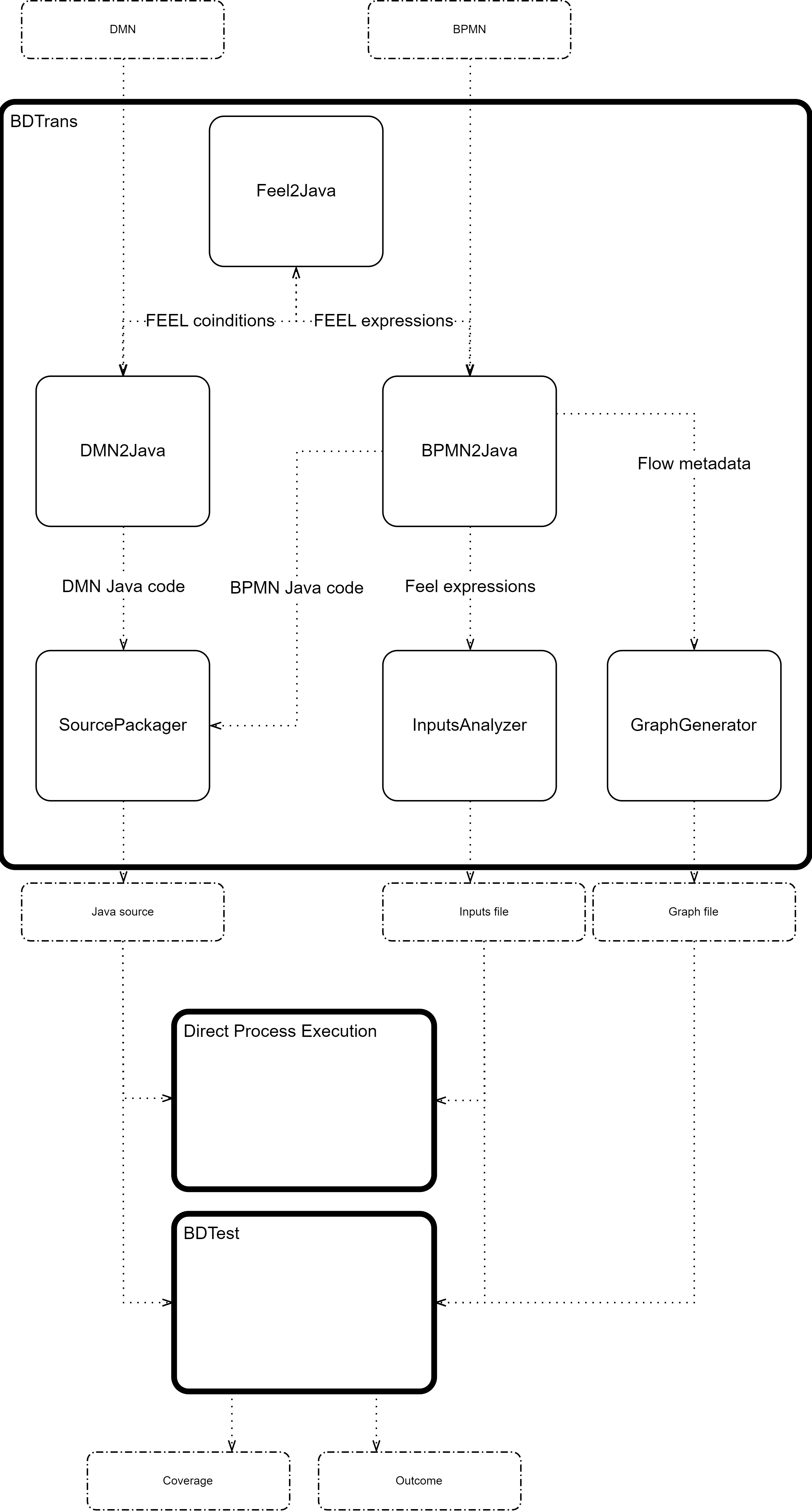}}}
\vspace*{8pt}
\caption{BDTransTest: overall proposed methodology breakdown.\label{overall_process}}
\end{figure}

\begin{enumerate}

\item We describe an automated translation from a \ac{BPMN}+\ac{DMN} process $({\cal B}, {\cal D})$ to a Java program $J({\cal B}, {\cal D})$.
  This allows both static inspection of $J({\cal B}, {\cal D})$ by an expert and framework-free execution of $({\cal B}, {\cal D})$ by using $J({\cal B}, {\cal D})$. We point out that the translation algorithm has been developed in a modular way, thus the usage of Java as the target language may be easily replaced by any other imperative programming language where all the \ac{BPMN} and \ac{DMN} features (most notably, parallelism) can be achieved or simulated.

  \item The translation described in the previous point also detects all {\em inputs} of $({\cal B}, {\cal D})$ with the corresponding domains. That is, we are able to detect which variables in the input business process ${\cal B}$ are expected to receive a value from the outside environment.

  \item For most business processes $({\cal B}, {\cal D})$, starting from the list of the inputs of $({\cal B}, {\cal D})$ from the previous point, we automatically synthesize and execute a testing plan for $({\cal B}, {\cal D})$. Such testing plan selects in a guided random way the values for each input and then executes $J({\cal B}, {\cal D})$ on such values. Our methodology allows the user to choose between different types of testing plans, e.g., with a fixed number of iterations or using statistical model checking. For business processes involving complex decisions, further disambiguating inputs from the user is required in order to perform the verification.

  \item We analyze the results of the testing phase described in the previous point, in order to compute the coverage achieved, both in terms of nodes and edges of  ${\cal B}$.

  \item We present the {\sf BDTransTest} tool, available at \github,
    which implements the methodology described above. An overview of the {\sf BDTransTest} tool is shown in Figure~\ref{overall_process}.

    \item Finally, we evaluate our {\sf BDTransTest} tool on \ac{BPMN}+\ac{DMN} processes from the literature.

\end{enumerate}

The rest of the paper is organized as follows. Section~\ref{related_work.sec} describes the state-of-the-art on \ac{BPMN}+\ac{DMN} execution and verification from the literature. Section~\ref{sec:background} provides the technological and theoretical background needed to understand this paper. 
Section~\ref{sec:tojava} describes in detail how the translation from a \ac{BPMN}+\ac{DMN} business process to a Java program is performed (corresponding to contributions 1 and 2 discussed above). Section~\ref{sec:analysis} shows how such translation is exploited to perform thorough testing of the input \ac{BPMN}+\ac{DMN} business process (corresponding to contributions 3 and 4 discussed above). Section~\ref{sec:expres} presents experimental results on using our approach on two case studies from the literature. Section~\ref{sec:conclusions} concludes the paper with final remarks and future developments.

%% file: related_work.tex
\section{Related Work}\label{related_work.sec}

Our main goal is verification of \ac{BPMN}+\ac{DMN} processes, thus we organize our literature review by first discussing the (not many) papers addressing the very same problem, then we also examine some works dealing with verification of \ac{BPMN} only and \ac{DMN} only. 

In~\citep{LFM21}, a process described by \ac{BPMN} and \ac{DMN} is translated into a Data Petri net. This allows to exploit verification techniques for Data Petri nets to verify \ac{BPMN} properties like data-aware soundness, as well as arbitrary temporal formulae expressing requirements on both the control-flow and data. In this work, we do not rely on an intermediate formalism like Data Petri nets, but we translate the \ac{BPMN}+\ac{DMN} process in an executable program in order to directly apply testing techniques. We also note that our approach does not have the limitations of~\cite{LFM21}, e.g., we are able to fully handle integer variables and arithmetic operations.
In~\citep{BS20}, the authors propose an extension of the \ac{BPMN} engine Camunda which allows to test processes described by \ac{BPMN} and \ac{DMN}, by also requiring in input a set of paths of the given \ac{BPMN} and a database of test cases to be customized. In our approach, we lift such limitations, by only requiring in input a \ac{BPMN}+\ac{DMN} process (possibly with disambiguations for complex \ac{DMN}s) and by only relying on a JVM for execution. 
In \citep{ARK19}, the reverse problem is considered of using \ac{BPMN}+\ac{DMN} to enable \ac{SysML} to perform verification, in the form of requirements traceability. Instead, in this paper we focus on \ac{BPMN}+\ac{DMN} verification.
Finally, in~\cite{KGR22}, a survey is presented that discusses the challenges that emerge when modeling, executing, and monitoring IoT-driven business rules using \ac{BPMN} 2.0 and \ac{DMN} standards.



As for literature on verifying a given \ac{BPMN} without \ac{DMN}, we may consider the following papers.
In~\cite{GHLW18}, a review of the current state-of-the-art on \ac{BPMN} direct execution (thus supporting also verification) is carried out, showing that very few vendors currently support the full \ac{BPMN} 2.0 standard.
In~\cite{FDAM12}, methodologies borrowed from Model-Driven Engineering are used to translate a \ac{BPMN} into a Petri Net, that is then executed and verified within the DENEB workflow.
In~\cite{PFP19}, a methodology is proposed to refactor a given \ac{BPMN}, in order to reduce quality faults, such
as unmeaningful elements, ﬁne-grained granularity or incompleteness.
In~\cite{CMRRT22}, a formalization and a tool are proposed for the animated execution of \ac{BPMN} collaborations.
In~\cite{VTS22}, a modelling approach is proposed that uses standard BPMN concepts to model \ac{IoT}-enhanced business processes without modifying its metamodel, so as to facilitate the execution of the \ac{IoT}-enhanced business processes independently from \ac{IoT} devices technology.
In~\cite{NQT25}, a methodology is proposed to find defects of flows in SAP business processes with a Machine-Learning approach.
In~\citep{IDDCG23}, a three-step approach is proposed that incrementally refines a
critical system specification, from a lightweight high-level model targeted to stakeholders, down to a
formal standard model that links requirements, processes and data.
In~\citep{AG16}, the given \ac{BPMN} is executed using existing frameworks, and an analysis of the resulting logs is performed in order to detect properties like absence of deadlocks and livelocks, as well as of dead elements.
In~\citep{BGB20,BGT20}, a tool is developed that covers three main aspects of verifying a \ac{BPMN}: a static verification, an interactive execution, and an automatic exploration of the possible scenarios.
In~\citep{KPS17}, a Web application is designed that
takes as input \ac{BPMN}-2.0-compliant business processes. The processes are ﬁrst translated into
model-speciﬁc veriﬁcation scripts in the SVL~\cite{GLMS13} verification language. Then, the verification
toolbox CADP~\cite{GLMS13} is
used to check either for functional properties. 
In~\citep{KDS14}, a methodology is proposed to build a knowledge-based for construction process reengineering by templates, rules and
patterns in order to verify, validate and modify configured process models. Petri Nets method is selected for the verification and
validation purpose.
In~\citep{LGT23}, the Business Process Evaluation and Research Framework for Enhancement and Continuous
Testing (bPERFECT) framework is proposed, which aims to guide business process testing research and
implementation. Secondary objectives include eliciting the existing types of testing, evaluating their
impact on efficiency and assessing the formal verification techniques that complement testing.
In~\citep{intrigila2021lightweight}, an approach is presented that allows stakeholders and software analysts to merge and
integrate behavioral and data properties in a \ac{BPMN} model. To this aim, a
lightweight \ac{BPMN} extension is proposed that specifically addresses the annotation of data properties in terms of
constraints, i.e., pre- and post-conditions that the different process activities must satisfy. 
In~\citep{KSRLK24}, a tool is designed that identifies control flow
errors in \ac{BPMN} models, make them understandable for modelers, and suggest corrections to resolve
them. 
In~\citep{NOAS16}, the input \ac{BPMN} is transformed to the
topological functioning model in order to check completeness of inputs, outputs and functioning
cycles of the entire specified system. The proposed approach is dedicated to the verification of the model at
the beginning of analysis, and it could be supplemented by other methods at the design stage.
In~\citep{FPPR12}, a Java based veriﬁcation approach is proposed
for Business Processes modeled using the \ac{BPMN} 2.0 standard.
The behavior of a set of interrelated
objects, corresponding to a \ac{BPMN} 2.0 speciﬁcation, are explored by means of a verification
algorithm using an ad-hoc unfolding technique.
In~\citep{Tak08}, it is shown how to  apply reachability and coverability analysis of Petri Nets to verification of business processes with
transactions and compensation.
In~\citep{CFPRTV21}, a verification approach is presented for \ac{BPMN} collaborations. It combines both
standard model checking techniques, through the MAUDE~\cite{BEM13} LTL model checker, and statistical model
checking techniques, through the statistical analyzer MultiVeStA~\cite{VGLC22}.
In~\citep{DVT19}, a
hierarchical verification technique is proposed for the state space analysis of a \ac{BPMN} based on a colored Petri Nets (CPN). A \ac{BPMN}
partitioning technique and rules for the transformation of a \ac{BPMN} into a CPN model are provided. The
partitioning approach supports the unstructured \ac{BPMN} design model, and the obtained CPN model also
supports hierarchical verification. 
In~\citep{SSM15}, 
a model checker based framework is presented to automate the veriﬁcation process of a \ac{BPMN}.
To this aim, the \ac{BPMN} is automatically converted
into PROMELA, the input language of
the SPIN model checker. Furthermore, the tool eases the
task of expressing Linear Temporal Logic correctness
requirements.
In~\citep{HBPQK22}, a formal semantics is proposed for a subset of \ac{BPMN}.
In contrast to transformational approaches, which give
a semantics to \ac{BPMN} by mapping it to some formal model (e.g., Petri nets), this
approach is based on a direct formalization in first-order logic that is then realized into the TLA$^+$ model checker input language.

Finally, some work only considers \ac{DMN} tables, without their \ac{BPMN}  environment. Namely, in~\cite{GCD21} a review is presented of tools allowing the verification of \ac{DMN} tables, while in~\cite{CKD24} a survey of the most common errors (e.g., missing rules) in modeling \ac{DMN} tables taken from large datasets is discussed.


Summing up, to the best of our knowledge, this paper is the first to propose an automatic approach to the translation, execution and testing of a business process described using \ac{BPMN} and also employing \ac{DMN} tables.

%% file: background.tex
\section{Background}\label{sec:background}

In this section, we provide the main notions needed to understand the rest of the paper, by describing the fundamentals of the \ac{FEEL} language (Section~\ref{feel_descr:subsec}), of \ac{BPMN} (Section~\ref{bpmn_descr:subsec}) and of \ac{DMN} (Section~\ref{dmn_descr:subsec}). 

\subsection{FEEL Language}\label{feel_descr:subsec}

\ac{FEEL} is a language designed to express simple expressions 
in a way that is easily understood by both professionals and developers \citep{feel}. It has been specified by the \ac{OMG} and can be used in both \ac{BPMN} and \ac{DMN} to write expressions where required. 

Namely, \ac{FEEL} 
allows to build intuitive expressions of the following types: 
    String,
    Numeric,
    List,
    Context,
    Temporal and
    Boolean. String type contains any sequence of characters between double quotes, also allowing concatenation between strings. Numeric type includes both integer and real numbers, and allows the standard mathematical operations (e.g., addition and multiplication) as well as some standard functions, e.g., {\tt abs}, {\tt floor}, {\tt sqrt}, etc. Temporal type allows to specify dates and times, and allow time-based arithmetical operations (e.g., {\tt time("08:00:00") + time("04:00:00")}). List type contains order sequences in square brackets, allowing 
    dereferentiation of one or multiple list elements (e.g., {\tt [1,2,3,4][item > 2]} returns {\tt [3,4]}, while {\tt [1,2,3,4][2]} returns {\tt 2}). 
    Context types are similar to Python dictionaries, e.g., {\tt \{a: "bye", b : 2\}}.
    Boolean expressions atomic proposition involve comparisons between the above defined types (e.g., {\tt v <= 3}, {s = "Hi"}, {\tt time("04:00:00") <= time("08:00:00")}), {\tt null} checks, type tests (e.g., {\tt 1 instance of string} is false, {\tt "1" instance of string} is true), list memberships (e.g., {\tt 1 in [1, 3, 5]} is true) and so on. Of course, the main Boolean operators (NOT, AND, OR) are also available. Finally, ranges may be build using the numeric and the temporal types, e.g., {\tt [3..10]}, also allowing operators to check for overlapping, inclusion and so on (e.g., {\tt overlaps before([1..5], [4..10])} is true).

\subsection{BPMN Notation}\label{bpmn_descr:subsec}


Figure~\ref{shipment.bpmn}, adapted from~\cite{LFM21} and that we use for reference in the following, shows an example of \ac{BPMN} business process. A \ac{BPMN} process is an annotated and labeled graph where nodes and edges may have different types, resembling a control flow diagram.
A number of {\em variables} may be used within each \ac{BPMN} process; usage of a variable include both simply reading its current value (e.g., as in the {\tt pLength = -1} check in Figure~\ref{shipment.bpmn}, that reads variable {\tt pLength}) or modifying its current value by writing a new value (e.g., as in the ``get length'' box in Figure~\ref{shipment.bpmn}, which writes a new value into variable {\\t pLength}, also see discussion below).
Namely, main node and edge types are the following (Section~\ref{subsec:limits} discusses the few \ac{BPMN} elements we do not consider here):
\begin{itemize}
\item start events, depicted as circle-shaped, are the entry points of the whole process, i.e., they have no ingoing edges and one outgoing edge; in Figure~\ref{shipment.bpmn}, the start event is the ``package received'' node;
  \item end events, depicted as bold-circle-shaped, are the exit points of the whole process (i.e., they have no outgoing edges and one ingoing edge) and are typically distinguished as correct end events (``ready for shipment'' in Figure~\ref{shipment.bpmn}) and error end events (also containing a bolt, see ``undefined length'', ``unsupported weight'' and ``no shipment'' end events in Figure~\ref{shipment.bpmn});
  \item tasks, depicted as rounded-edges rectangles and having exactly one ingoing and one outgoing edge, describe some action to be taken, such as the ``get length'', ``measure weight'', ``determine mode'' and ``choose consent'' tasks in Figure~\ref{shipment.bpmn}.
    Several types of task exist. Here, we focus on those which may change the value of a variable, i.e., {\em user tasks} (depicted with an avatar figure, see ``measure weight'' task in Figure~\ref{shipment.bpmn}), {\em manual tasks}, {\em script tasks}, {\em service tasks} and {\em business rule tasks} (depicted with a table as they are linked to a \ac{DMN} table, see Section~\ref{dmn_descr:subsec}). Namely, user tasks and manual tasks may assign to a variable the value decided from an external user. Instead, script, service and business rule tasks may assign to a variable the value of a given \ac{FEEL} expression and of a \ac{DMN} table, respectively. We will also consider {\em send tasks} and {\em receive tasks}, which allow to send and receive messages when a \ac{BPMN} process is divided in concurrent threads of execution;
  \item gateways, depicted as a diamonds, represents points in which some choice must be done; they can be further divided as follows:
    \begin{itemize}
    \item {\em exclusive gateways}, where the diamond contains a cross, have one incoming and multiple outgoing edges. Such outgoing edges are labeled with \ac{FEEL} Boolean expression, so that the first edge evaluating to true is executed. There are 4 of such gateways in Figure~\ref{shipment.bpmn}, where a crossed edge identifies the default case;
      \item {\em parallel gateways}, where the diamond contains a plus sign, and {\em inclusive gateways}, where the diamond contains a circle, both have one incoming and multiple outgoing edges. Inclusive gateways outgoing edges again have Boolean \ac{FEEL} conditions, while parallel gateways does not. Differently from the preceding case, all edges (for parallel gateways) and all edges evaluated to true (for inclusive gateways) are {\em concurrently} executed; 
    \item  {\em joining gateways} have multiple incoming and one outgoing edges, and are meant for re-join of different path which were split before (the symbol inside the diamond is either a cross, a plus or a circle, depending on what it is re-joining): there is one such gateway in Figure~\ref{shipment.bpmn};
    \end{itemize}
  \item annotations, depicted as rectangles and labeled with a variable, indicate which task or event read and/or writes the corresponding variable: outgoing dotted edges to tasks represent a read access, ingoing dotted edges from tasks represent a write operation (there are 5 such annotations in Figure~\ref{shipment.bpmn}).
\end{itemize}

\begin{figure}
\centering\framebox{
\begin{minipage}{0.5\hsize}        
    \scriptsize
    \verbatiminput{shipment/shipment.graph}   
\end{minipage}%
\begin{minipage}{0.5\hsize}
    \centerline{\includegraphics[width=\hsize]{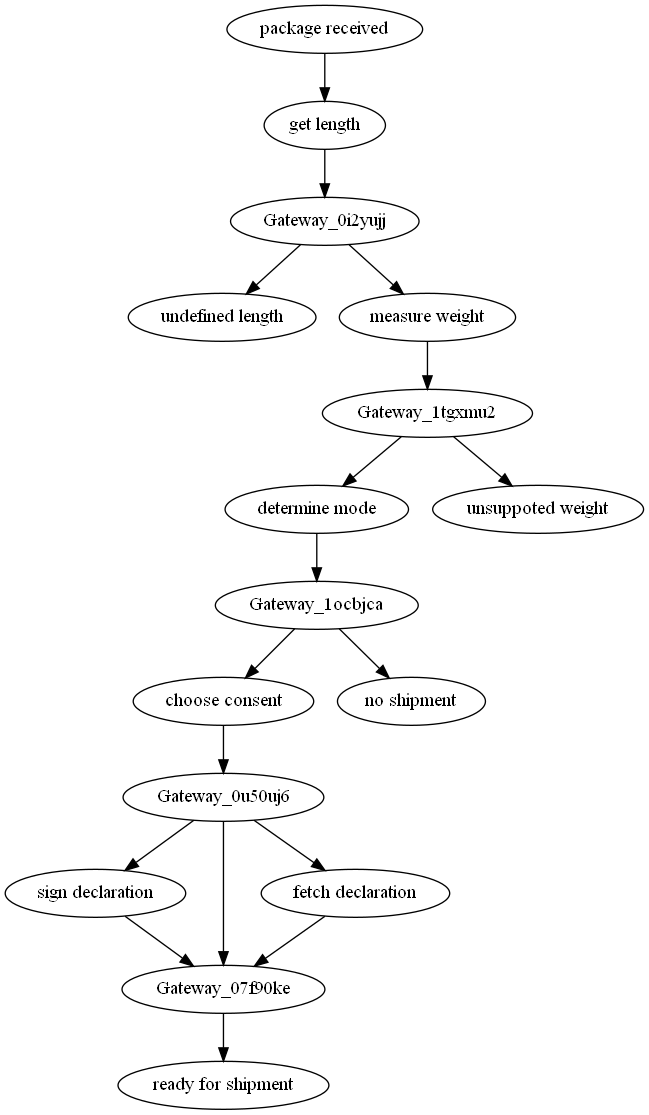}}
\end{minipage}}
\vspace*{8pt}
\caption{Process graph data from the \Example{Shipment} business process and corresponding graphical representation.\label{shipment.graph}}
\end{figure}

\begin{figure}
\centerline{\includegraphics[width=0.8\hsize]{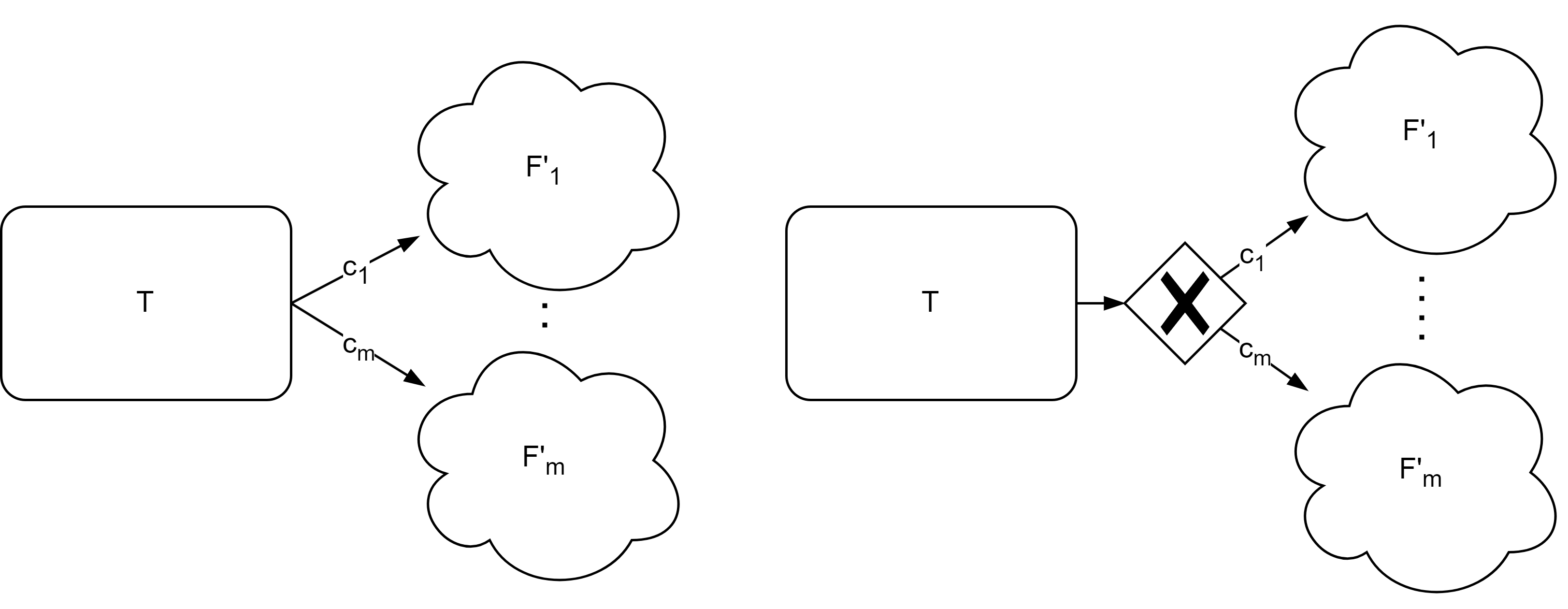}}
\vspace*{8pt}
\caption{Fixing a multiple-output task using a gateway.\label{generic_fix.bpmn}}
\end{figure}

As for the \ac{BPMN} representation as a file, in this paper we will refer to the XML encoding used by many \ac{BPMN} editors like Camunda. Such encoding uses a predefined XML schema to represent all \ac{BPMN} elements (tasks, events, gateways, annotations for variables...) listed above.
Furthermore, we assume to have a function \textsl{extractGraph} which takes a \ac{BPMN} representation and returns the underlying graph, in which annotations are discarded and node/edge types are not distinguished (see Figure~\ref{shipment.graph} for the graph extracted from Figure~\ref{shipment.bpmn}). Finally, we note that \ac{BPMN} also allows tasks with more than one outgoing edge; in the following, we suppose that \ac{BPMN} processes are preprocessed so as to add exclusive gateways after such tasks as shown in Figure~\ref{generic_fix.bpmn} (such preprocessing is actually performed also in our implementation tool \BDTransTestNoSp).

\subsubsection{Variables in \ac{BPMN} Processes}\label{vars_in_bpmn.subsec}

\begin{figure}
\centerline{\includegraphics[height=0.6\textheight]{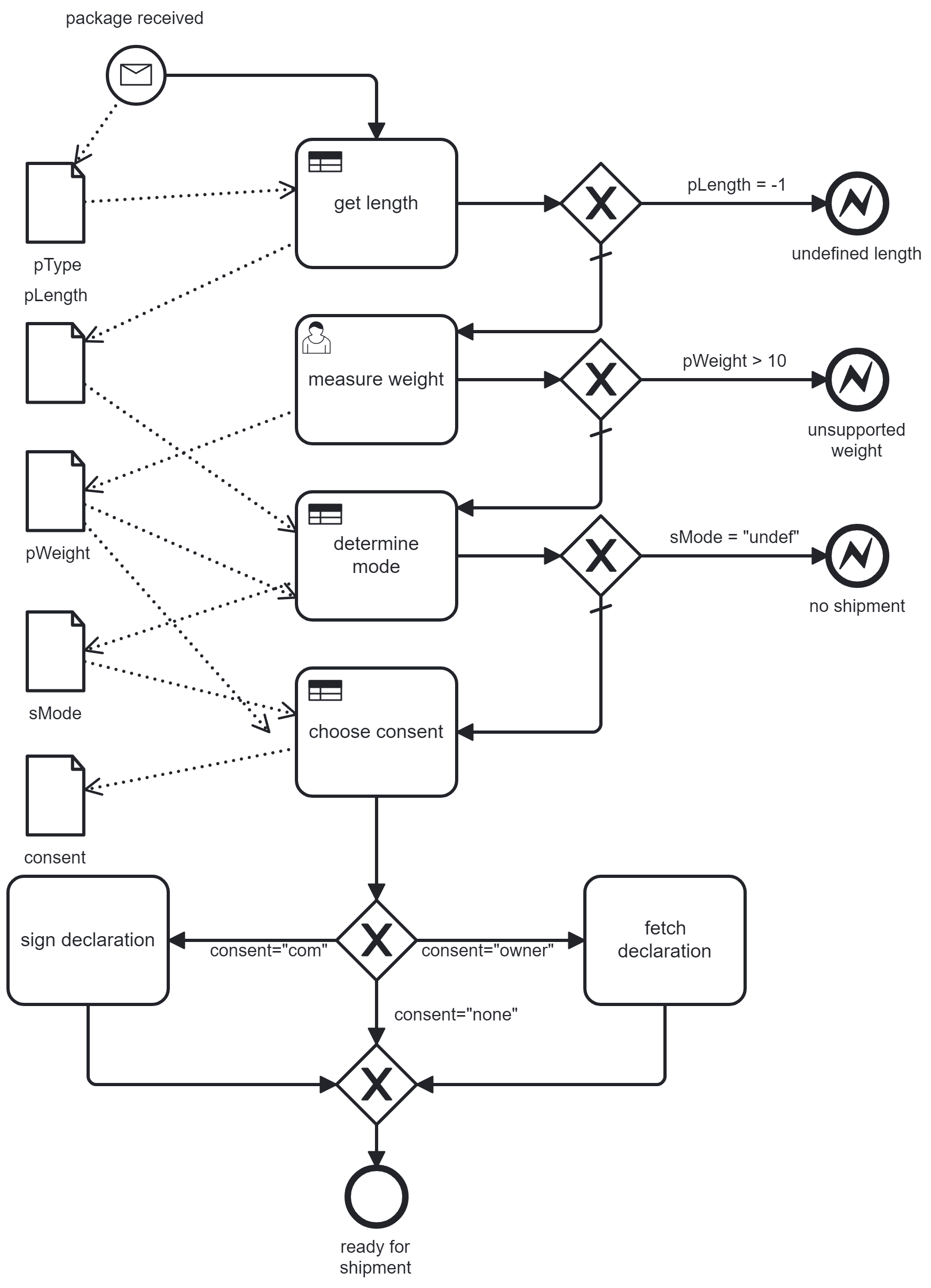}}
\vspace*{8pt}
\caption{\textit{Shipment} Business Process from~\cite{LFM21}.\label{shipment.bpmn}}
\end{figure}

As discussed above, user and manual tasks, as well as start events, may involve a write operation on a variable which is decided by an external user, while script and business rule tasks may write a variable basing on the \ac{BPMN} process itself. To highlight such behaviour, we will call {\em input variables} those which are written by start events, user tasks or manual tasks, while we call {\em process variables} the other ones.
We note that, strictly speaking, a variable in a \ac{BPMN} process may be both an input variable and a process variable, e.g., if it is written both by the start event and by a business process rule (e.g., suppose that ``get length'' business rule task in Figure~\ref{shipment.bpmn} writes {\tt pType} again, instead of {\tt pLength}). This however entails a re-usage of variables which could hamper \ac{BPMN} readability and lead to programming errors, as it happens with variables re-usage in high-level programming languages. For this reason, we pre-process the input \ac{BPMN} process to detect variables which are both input and process variables, so that the process engineer may re-name such variables. We also note that such pre-process may be valuable to find some early errors in the \ac{BPMN} process.

Furthermore, if a \ac{BPMN} process is drawn at an early process development stage, it may not correctly specify write operations on variables. As an example, in a first draft for the \Example{Shipment} \ac{BPMN} process in Figure~\ref{shipment.bpmn}, the annotation for {\tt pType}  may be omitted, and variable {\tt pType} may be directly read inside the \ac{DMN} ``get length''. In order to capture this behaviour, also such only-read-and-never-written variables are considered input variables.

\subsection{DMN Notation}\label{dmn_descr:subsec}

\begin{figure}
\centerline{\includegraphics[width=0.9\hsize]{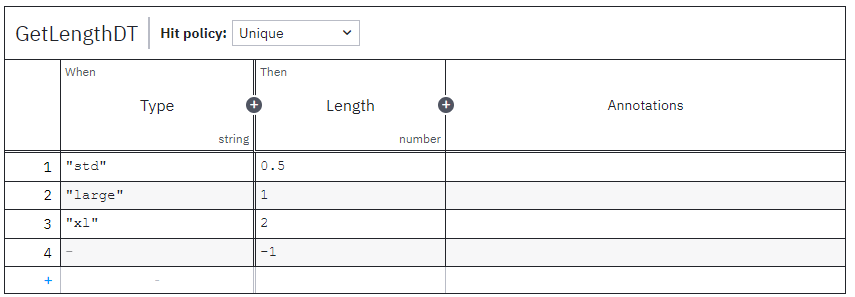}}
\vspace*{8pt}
\caption{\textit{Get Length} Decision Table used in the \textit{Shipment} Business Process from~\cite{LFM21}.\label{get_length.dmn}}
\end{figure}
\begin{figure}
\centerline{\includegraphics[width=0.9\hsize]{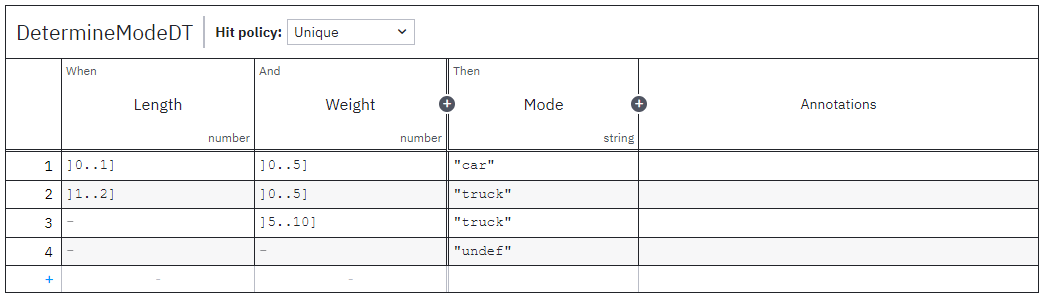}}
\vspace*{8pt}
\caption{\textit{Determine Mode} Decision Table used in the \textit{Shipment} Business Process from~\cite{LFM21}.\label{determine_mode.dmn}}
\end{figure}
\begin{figure}
\centerline{\includegraphics[width=0.9\hsize]{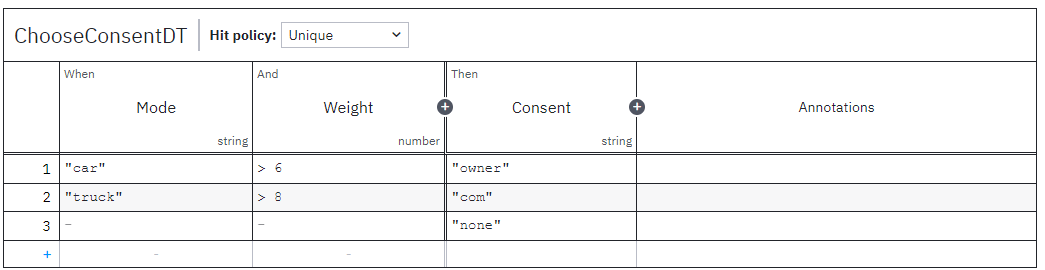}}
\vspace*{8pt}
\caption{\textit{Choose Consent} Decision Table used in the \textit{Shipment} Business Process from~\cite{LFM21}.\label{choose_consent.dmn}}
\end{figure}

Figures~\ref{get_length.dmn}--\ref{choose_consent.dmn}, again adapted from~\cite{LFM21}, show 3 examples of \ac{DMN} tables used within a \ac{BPMN} business process (the same depicted in Figure~\ref{shipment.bpmn}). 
Each of such tables is linked with a \ac{BPMN} task; a task embodying a \ac{DMN} table is called a {\em business rule task}. As an example, the \ac{DMN} table in Figure~\ref{get_length.dmn} is linked to the ``get length'' business rule task in Figure~\ref{shipment.bpmn}, as well as \ac{DMN} tables in Figures~\ref{determine_mode.dmn}~and~\ref{choose_consent.dmn} are linked to business rule tasks ``determine mode'' and ``choose consent'', respectively. The link between a \ac{DMN} table and the corresponding business rule task also specifies how arguments are passed, i.e., how the \ac{BPMN} process variables maps into the \ac{DMN} variables or expressions used in the header row.

Figure~\ref{generic.dmn} shows the generic form of a \ac{DMN} table. Namely, such table is composed, from left to right, by at least one {\em input} column and at least one {\em output} column, with a double vertical line separating inputs and outputs (in some editors, also different colors are used). A final column may be used to hold some annotations. The first (header) row of a \ac{DMN} table labels each input/output column $c$ with a variable $v(c)$ occurring in the corresponding \ac{BPMN}. 
The other rows contain, in the input part, either constant values or ranges in the domain of the corresponding \ac{FEEL} expression, or a fixed value in the domain of the corresponding output column variable. Namely, each row represents a conditional assigment to all output column variables, depending on the input columns to fulfill the corresponding \ac{FEEL} conditions shown in the same row. In the following, for the sake of clarity, we assume to employ the {\em first policy}, i.e., if more than one row is satisfied by the current input variables values, the first row is selected. However, our \BDTransTest tool is also able to handle other policies, i.e., the ``unique'' policy and the ``any'' policy.

In Figure~\ref{generic.dmn}, the first group of columns refer to the table input variables ($in_1,\ldots, in_k$), the second group to output variables ($out_1, \ldots, out_n$), and $c_{i, j}$ is the content, as a \ac{FEEL} expression, of \ac{DMN} cell $(i, j)$. To define the formal semantics of such generic \ac{DMN} table, let $\hat{c}_{i, j}$ be true iff the value of input variable $v$ corresponding to input column $j$  satisfies $c_{i, j}$. Since $c_{i, j}$ may contain either a value $\bar{v}$ or a range $[a..b]$, $\bar{c}_{i, j}$ is true if $v = \bar{v}$ in the first case, and if $v \in [a, b]$ in the second case. Note that, in the input columns, a dash means ``don't care'', in which case $\bar{c}_{i, j}$ is always true. Given this, the semantics of such table is an AND between implications involving input and output variables of each row. More in detail, the semantics of the generic \ac{DMN} table in Figure~\ref{generic.dmn} is the Boolean formula:

$$\land_{i = 1}^{m} ((\land_{j=1}^k \bar{c}_{i, j} \land \land_{i' = 1}^{i - 1}\neg(\land_{j=1}^k \bar{c}_{i', j})) \to (\land_{j=1}^n (out_j = v_{i, j}))) \land$$ $$(\land_{i = 1}^{m}\neg(\land_{j=1}^k \bar{c}_{i, j}) \to (\land_{j=1}^n (out_j = v_{m+1, j})))$$

\begin{figure}
\centering
        \begin{tabular}{|c|c|c||c|c|c|}\hline
            \textbf{$in_1$} & \ldots & \textbf{$in_k$}  & \textbf{$out_1$} & \ldots & \textbf{$out_n$} \\\hline
            $c_{1,1}(in_1)$ & \ldots & $c_{1,k}(in_k)$ & $v_{1,1}$ & \ldots & $v_{1,n}$ \\\hline            
             \vdots  &  \vdots &  \vdots&  \vdots&  \vdots&  \vdots\\\hline
             $c_{m,1}(in_1)$ & \ldots & $c_{m,k}(in_k)$ & $v_{m,1}$ & \ldots & $v_{m,n}$ \\\hline            
             - & - & - & $v_{m+1,1}$ & \ldots & $v_{m+1,n}$ \\\hline
        \end{tabular}       
\vspace*{8pt}
\caption{A generic Decision Table \texttt{GenTabDT}.\label{generic.dmn}}
\end{figure}

%% file: translation.tex
\section{From BPMN and DMN to (Java) Code} \label{sec:tojava}

In this section, we describe the first part of our \BDTransTest tool, i.e., the algorithm which takes as input a \ac{BPMN} process (possibly referencing \ac{DMN} decision tables) and outputs a set of Java classes. By executing the main of such classes with a given assignment to \ac{BPMN} variables, the same trace is returned which results from executing the input \ac{BPMN}. From now on, we will refer to such algorithm as \BDTransNoSp, as depicted in Figure~\ref{overall_process}.

The rest of this section is organized as follows. Section~\ref{building:subsec} introduces the main auxiliary routines and libraries used by \BDTransNoSp. Section~\ref{overall:subsec} explains the overall high-level organization of \BDTransNoSp. Section~\ref{bpmn_trans:subsec} details the translation of the main tasks, events and gateways of a \ac{BPMN}. Section~\ref{dmn_trans:subsec} shows how \ac{DMN} tables are translated, and Section~\ref{sec:tojava-generation-business} shows how they are linked to the main \ac{BPMN}.
Section~\ref{sec:tojava-generation-messages} deals with \ac{BPMN} tasks containing send/receive message operations.
Section~\ref{sec:tojava-example} shows an example of a complete translation of \ac{BPMN} process into a Java code. 
Section~\ref{implementation:subsec} discusses the main implementation details of the translation inside our \BDTransTest tool.
Section~\ref{limitations:subsec} discusses the current limitations of our approach. 
Finally, Section~\ref{proof:subsec} shows that \BDTrans behaves correctly.


\subsection{\BDTrans Auxiliary Functions and Libraries}\label{building:subsec}

The Java classes output by \BDTrans rely on an execution framework, called \FunctionDef{BPMNExecutionLayer} (\BeL for short in the following figures), which hides all language-specific library calls and structures, leaving the generated code as clean and easy to read as possible. As an example, such framework includes classes to maintain the current status of a process execution, to retrieve the input values for \ac{BPMN} variables, to output the execution trace, to perform comparisons between \ac{BPMN} variables and so on.

Furthermore, \BDTrans also includes an auxiliary subroutine \FEELJava that is able to translate a generic \ac{FEEL} expression to Java code, by also relying on the \BeL framework. Namely, given a \ac{FEEL} expression $e$, \FEELJava first considers the abstract parse tree of $e$. Then, the translation is performed top-down, from the root to the leaves, by using the most suitable Java type for each FEEL type (e.g., the Boolean FEEL type is cast into the {\tt boolean} Java type) and the most suitable Java operator for each FEEL operator (e.g., the FEEL plus sign can be translated in the Java plus sign both for numbers and strings). This also entails calling Java functions when needed, e.g., to translate the {\tt abs} or the {\tt length} FEEL functions, or to map the equality sign to {\tt equals} Java function. 

\subsection{Overall \BDTrans Organization}\label{overall:subsec}

In the following, we will refer to the input of \BDTrans as a pair $({\cal B}, {\cal D})$, where ${\cal B}$ is a \ac{BPMN} and ${\cal D}$ is a (possibly empty) set of decision tables linked to ${\cal B}$.

Given this, the overall structure of \BDTrans is described in Algorithm~\ref{main_algo}. Namely,
function \BDTrans first analyzes all annotations and all \ac{FEEL} expressions used in the \ac{BPMN}, i.e., in tasks, gateway conditions, and decision table rules, and derives the \textit{process} and {\em input variables} of ${\cal B}, {\cal D}$ (line~\ref{get_vars.step}, see also Section~\ref{vars_in_bpmn.subsec}). 
As an example, for the \Example{Shipment} \ac{BPMN} process in Figure~\ref{shipment.bpmn}, the following variables are identified: {\tt pType, pWeight, consent, pLength, sMode}. Of such variables, {\tt pType} and {\tt pWeight} are the input variables, as they are written by the start event and the ``measure weight'' user task, respectively; the other ones are process variables. Furthermore, for each of such variable, the corresponding domain is also derived. Namely, comparisons with fixed \ac{FEEL} constants are used to derive the correct Java type to be used; e.g, if the  \ac{FEEL} expression {\tt v < 3.5} occurs somewhere in the input \ac{BPMN}, then variable {\tt v} will be declared as {\tt Double}.
Both variable names and their Java types are collected in the returned sets $V_I$ (for input variables) and $V_P$ (for process variables), which are then used by the routines for \ac{DMN} and \ac{BPMN} translation. Namely, both $V_I$ and $V_P$ contains triplets $(v, t, d)$, where $v$ is a variable, $t$ its Java type ({\tt Integer, Double, String} etc.) and $d$ contains information needed by \BDTestNoSp, thus it will be explained in Section~\ref{sec:tojava-inputs}.

After this, \BDTrans detects all \ac{DMN} tables in the input \ac{DMN} ${\cal D}$, and separately translates each of them into three Java classes via function \FunctionDef{translateDMN} (line~\ref{dmn_start.step}). 
Formal details of such translation are provided in Section~\ref{dmn_trans:subsec}. Then, a further Java class is generated by auxiliary function \FunctionDef{translateBPMN} (detailed in Algorithm~\ref{aux_algo} and discussed in Section~\ref{bpmn_trans:subsec}) for the input \ac{BPMN} ${\cal B}$
in line~\ref{bpmn_trans.step}.
Finally, such class is returned in line~\ref{bpmn_end.step}, together with the other two outputs shown in Figure~\ref{overall_process}, i.e., the set of input variables $V_I$ returned by \FunctionDef{getVarsDoms}, and the underlying graph of ${\cal B}$ (obtained by invoking \FunctionDef{extractGraph}, already mentioned in Section~\ref{bpmn_descr:subsec}).

\begin{algorithm}
  \caption{Main Algorithm for \BDTrans}\label{main_algo}
  \begin{algorithmic}[1]
    \Function{BDTrans}{BPMN process ${\cal B}$, DMN tables set ${\cal D}$}
    \State $\langle V_I, V_P\rangle$ $\gets$ getVarsDoms(${\cal B}$, ${\cal D}$) \label{get_vars.step}
  \State $P$ $\gets$ translateDMN(${\cal D}, V_I \cup V_P$) \Comment{see Section~\ref{sec:tojava-generation-dmn}}
  \label{dmn_end.step}\label{dmn_start.step}
\State $P$ $\gets$ $P$ $\cup$ translateBPMN($P$, ${\cal B}$, ${\cal D}$, $V_I$, $V_P$) \label{bpmn_trans.step}\Comment{see Section~\ref{sec:tojava-generation-flow}} 
  
\State {\bf return} $\langle P, V_I, $extractGraph$({\cal B})\rangle$\label{bpmn_end.step}
\EndFunction
\end{algorithmic}
\end{algorithm}

\begin{algorithm}
  \caption{Auxiliary algorithm for \BDTrans}\label{aux_algo}
  \begin{algorithmic}[1]
    \Function{translateBPMN}{Output file $P$, BPMN process ${\cal B}$, DMN tables set ${\cal D}$, variables and domains $V_I, V_P$}
    \State $P$ $\gets$ $P$ $\cup$ TranslateJava($V_I, V_P$) \label{declare_vars.step}
    \State $P$ $\gets$ $P$ $\cup$ fixed Java code for initialization \label{init_java.step}
    \For{BPMN element $u$ in ${\cal B}$}\label{aux_forall.step}
    \If{$u$ is the Start Event, an End Event or a generic Task}
    \State $P$ $\gets$ $P$ $\cup$ translateSET($u, V_I, V_P$) \Comment{see Section~\ref{sec:tojava-generation-flow}}
    \ElsIf{$u$ is a Gateway}
    \State $P$ $\gets$ $P$ $\cup$ translateG($u, V_I \cup V_P$) \Comment{see Section~\ref{sec:tojava-generation-gateways}}\label{translateG.step}
    \ElsIf{$u$ is a Business Rule Task}
    \State $P$ $\gets$ $P$ $\cup$ translateB($u, V_I \cup V_P$) \Comment{see Section~\ref{sec:tojava-generation-business}}
    \EndIf
    \EndFor \label{aux_forall_end.step}
    \State {\bf return} $P$
    \EndFunction
  \end{algorithmic}
\end{algorithm}

\subsection{Overall \ac{BPMN} Translation} \label{bpmn_trans:subsec}\label{sec:tojava-generation-flow}

To better illustrate how function \FunctionDef{translateBPMN} in Algorithm~\ref{aux_algo} works, consider the generic overall \ac{BPMN} process \FunctionDef{GenericBPMN} in Figure~\ref{generic.bpmn}.
Note that, in the following figures, for sake of generality we shall indicate with a cloud symbol (which cannot be confused with an actual \ac{BPMN} element) a generic sub-diagram $F$ such that its last element is not an exclusive, a parallel or an inclusive gateway. 
We also denote the starting and ending \ac{BPMN} tasks of $F$ with $First(F)$ and $Last(F)$, respectively.
Therefore, the \FunctionDef{GenericBPMN} process in Figure~\ref{generic.bpmn} starts with event \textit{S}, continues with a sub-flow \textit{F} and ends with the end events $E_1, \ldots, E_n$, which may be success (without loss of generality, the first $k$) or error end events  (without loss of generality, the other $n - k$).

Given this, function \FunctionDef{translateBPMN} synthesizes the single Java class $J({\cal B})$ to be output for the input \ac{BPMN} process ${\cal B}$ as follows. The class name for  $J({\cal B})$ coincides with that of ${\cal B}$.
All variables detected in the previous phase, and passed to \FunctionDef{translateBPMN} as arguments $V_I, V_P$, are declared as class members with the generic Java {\tt Object} type (line~\ref{declare_vars.step} of Algorithm~\ref{aux_algo}). When a task or a start event will need to assign a specific value to a variable, a cast is performed using suitable types from \BeL.
In this starting phase, also 3 member methods are added for initialization and overall execution: {\tt init, execute, main} (line~\ref{init_java.step} of Algorithm~\ref{aux_algo}). $J({\cal B})$ for the \ac{BPMN} process in Figure~\ref{generic.bpmn} is shown in Figure~\ref{generic.bpmn.java}. 

After such declarations, 
each \ac{BPMN} element 
in ${\cal B}$
is translated in a corresponding Java method class inside $J({\cal B})$. Such translation depends on the \ac{BPMN} element being a start event, an end event, a task or a gateway (lines~\ref{aux_forall.step}--\ref{aux_forall_end.step} of Algorithm~\ref{aux_algo}). Each method follows the $t$\_$N$ naming scheme, being $t$ the type of element modeled (\texttt{EVENT}, \texttt{TASK}, or \texttt{GATEWAY}) and $N$ the element unique identifier in the \ac{BPMN} process. If present, the element description is also appended to the method name, in order to make it more human-understandable.

As for the method body, first of all we note that each edge between two \ac{BPMN} elements ($e_1$, $e_2$) is translated so that the Java method corresponding to $e_1$ invokes the method corresponding to $e_2$. The rationale behind this translation is to maintain a 1:1 relationship between the generated code and the \ac{BPMN} elements, in order to allow an easy traceability of both the code and its execution trace with respect to its defining elements.

For start events and tasks, if they are not writing any variable, the method body only contains a call to the method corresponding to the successor \ac{BPMN} element, as described above. Such a call also takes as argument the current task source, so that execution traces can be otput in the verification phase. For send and receive tasks, suitable \BeL methods are invoked to manage the message exchange; see Section~\ref{sec:tojava-generation-messages} for a more detailed description. If a write operation is performed on a process variable in $V_P$ by a script, service or business rule task, then the corresponding variable is accordingly set by using \FEELJava or by evaluating the corresponding \ac{DMN} table (see Section~\ref{sec:tojava-generation-business}), respectively. If instead the start event or a user/manual task modify an input variable in $v \in V_I$, then we assume a list of input values $\ell(v) = \langle w_1, \ldots, w_k\rangle$ is provided for variable $v$; Section~\ref{sec:analysis} describes how this is actually performed. Exploiting this fact, the method body for the start event of user/manual task uses \BeL to keep track of how many values $j$ have already been used (at the start, $j=0$), and to select $v_{j + 1}$ as the new value for $v$. If $j\geq k$, then $w_k$ is used. Such lists of input values must be provided either by the user of \BDTransTestNoSp, or it is automatically generated by \BDTransTest itself (see Section~\ref{sec:analysis}). For end events, the method body only contains a call to {\tt BeL.success} or {\tt BeL.error}, depending on the type of the end event. This corresponds to the output of function \FunctionDef{translateSET} in Algorithm~\ref{aux_algo} (also see Figure~\ref{generic.bpmn.java}). Note that function \FunctionDef{translateSET} needs to distinguish between input variables $V_I$ and process variables $V_P$, while functions \FunctionDef{translateG} and \FunctionDef{translateB} do not.

As for gateways, the method body is more complex and is described in Section~\ref{sec:tojava-generation-gateways}.

\begin{figure}
\centerline{\includegraphics[height=0.2\textheight]{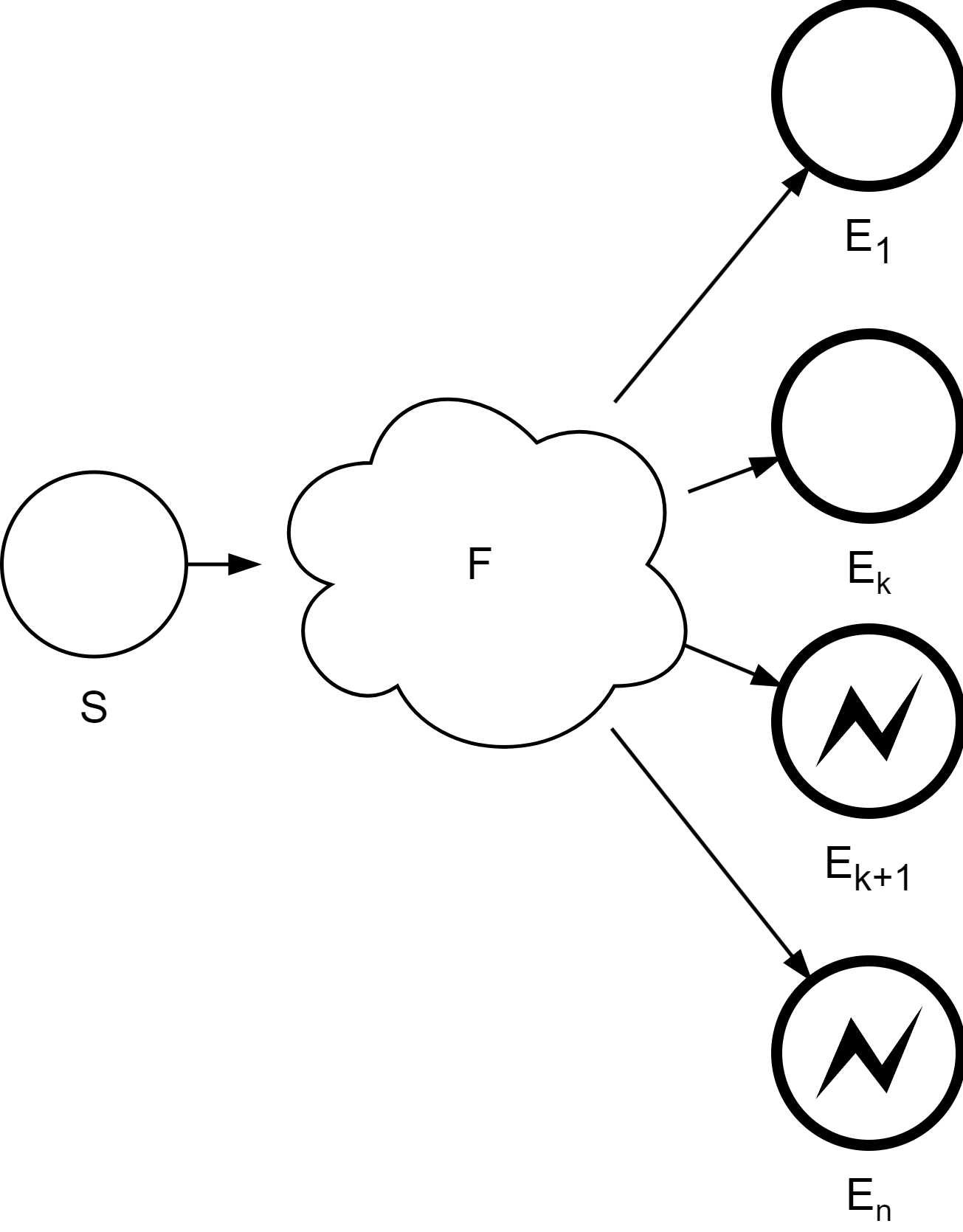}}
\vspace*{8pt}
\caption{A generic \ac{BPMN} Process \Example{GenericBPMN}.\label{generic.bpmn}}
\end{figure}

\begin{figure}
\begin{lstlisting}[language=Java,frame=tlbr]
class GenericBPMN { 
 // declarations of all process variables as class members
 public void init() { /* process variables init */ }
 public void execute() {
  BeL.executeProcess("GenericBPMN",this::init,this::EVENT_S);
 }
 public static void main(String[] args) {
  GenericBPMN process = new GenericBPMN();
  process.execute();
 }
 public void EVENT_S(BeL.ProcessStatus s) {
  TASK_$First(F)$(s.withCurrent("S"));
 } 
 $\ldots$
 public void TASK_$T$(BeL.ProcessStatus s) {
  // $T'$ $:=$ successor of $T$ with type $t$
  $t$_$T'$(s.withCurrent($T$));
 }
 $\ldots$
 public void EVENT_E$_1$(BeL.ProcessStatus s) { BeL.success(s); }
 $\ldots$
 public void EVENT_E$_k$(BeL.ProcessStatus s) { BeL.success(s); }
 public void EVENT_E$_{k+1}$(BeL.ProcessStatus s) { BeL.error(s); }
 $\ldots$
 public void EVENT_E$n$(BeL.ProcessStatus s) { BeL.error(s); }
}
\end{lstlisting}
\vspace*{8pt}
\caption{Java code generated for the generic \ac{BPMN} Process in Figure \ref{generic.bpmn}.}\label{generic.bpmn.java}
\end{figure} 


\subsubsection{Translation of Gateways} \label{sec:tojava-generation-gateways}

\begin{figure}
    \centerline{\includegraphics[width=0.9\hsize]{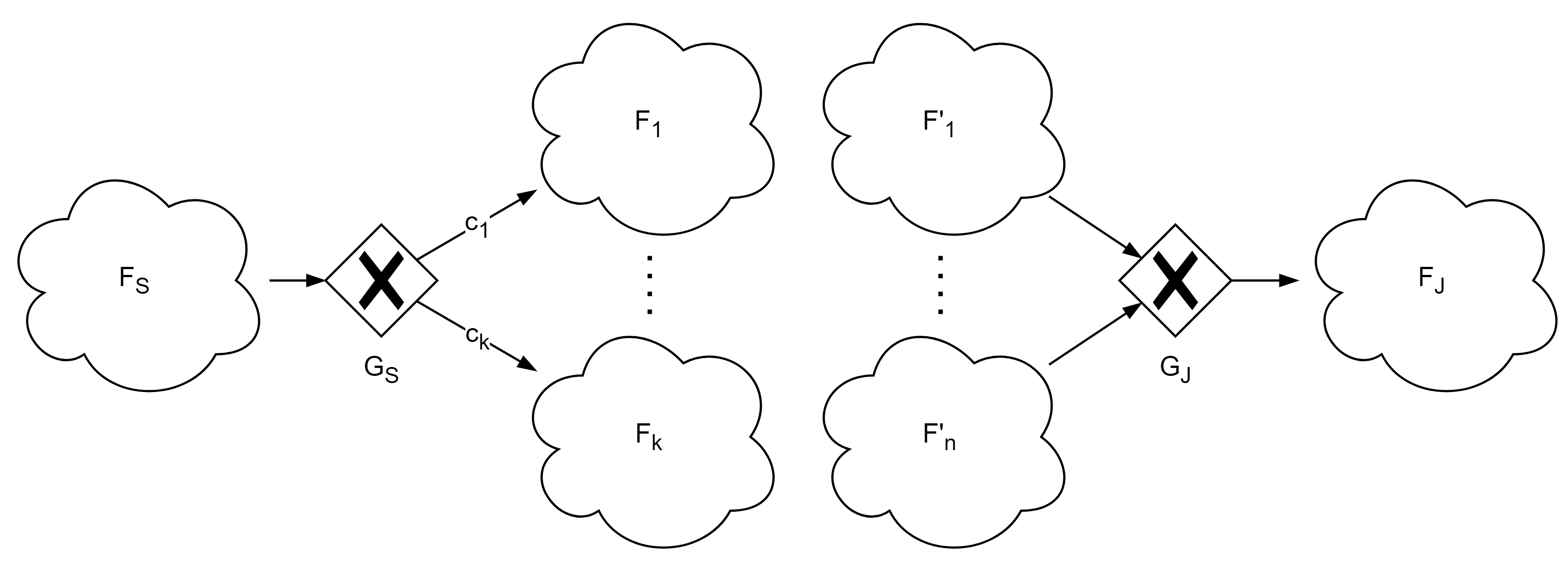}}
\vspace*{8pt}
\caption{A generic \ac{BPMN} Process \Example{GenericGW} with gateways.\label{generic_gateway.bpmn}}
\end{figure} 

To better illustrate how \ac{BPMN} gateways are translated by function \FunctionDef{translateG} (line~\ref{translateG.step} of Algorithm~\ref{aux_algo}), consider the generic overall \ac{BPMN} process \FunctionDef{GenericGW} in Figure~\ref{generic_gateway.bpmn}, where both an exclusive gateway $G_S$ and a joining gateway $G_J$ are employed.
Namely, $G_S$ has one incoming edge (from $F_S$) and $k$ outgoing edges targeting $F_1,\ldots,F_k$ with \ac{FEEL} conditions $c_1,\ldots,c_k$, respectively, whereas $G_J$ has $n$ incoming edges from \ac{BPMN} sub-processes $F'_1,\ldots,F'_n$  and a single outgoing edge targeting $F_J$.

\begin{figure}
\begin{lstlisting}[language=Java,frame=tlbr]
 $\ldots$
 public void $t_S$_$Last(F_S)$(BeL.ProcessStatus s) {
  // $t_S$ is the type of the $Last(F_S)$
  GATEWAY_$G_S$(s.withCurrent($Last(F_S)$));
 }
 public void GATEWAY_$G_S$(BeL.ProcessStatus s) {
  // for all $i$, $t_i$ is the type of the $First(F_i)$
  if ($\hat{c}_1$){
   $t_1$_$First(F_1)$(s.withCurrent("$G_S$"));
  } 
  $\ldots$  
  else if ($\hat{c}_k$){
   $t_k$_$First(F_k)$(s.withCurrent("$G_S$"));
  } else { 
    BeL.noDefaultCaseError(s); 
  }
 }
 $\ldots$
 public void $t'_1$_$Last(F'_1)$(BeL.ProcessStatus s) {
  // for all $i$, $t'_i$ is the type of the $Last(F'_i)$
  GATEWAY_$G_J$(s.withCurrent($Last(F'_1)$));
 }
 $\ldots$
 public void $t'_n$_$Last(F'_n)$(BeL.ProcessStatus s) {
  GATEWAY_$G_J$(s.withCurrent($Last(F'_n)$));
 }
 public void GATEWAY_$G_J$(BeL.ProcessStatus s) {
  // $t_J$ is the type of the $First(F_J)$
  $t_J$_$First(F_J)$(s.withCurrent("$G_J$"));
 }
 $\ldots$
\end{lstlisting}
\vspace*{8pt}
\caption{Java code generated for the \ac{BPMN} Process in Figure \ref{generic_gateway.bpmn}.}\label{generic_gateway.bpmn.java}
\end{figure} 

The code generated for the two subprocesses of Figure~\ref{generic_gateway.bpmn} is shown in Figure~\ref{generic_gateway.bpmn.java}, where the each outgoing edge condition $c_i$ of the exclusive gateway $G_S$ has been translated by \FEELJava into guard $\hat{c}_i$ of an \texttt{if} statement, whose body calls the method corresponding to the edge $c_i$ target. Note that, if an exclusive gateway has no \textit{default branch}, the translator automatically adds a last \texttt{else} statement calling an internal method that ends the process with a failure. In this way, capturing design flaws such as unhandled conditions becomes easier for the business process developers and for the verification tools.

Joining gateways, such as $G_J$ in Figure \ref{generic_gateway.bpmn}, are translated as edges between standard \ac{BPMN} elements (see Section \ref{sec:tojava-generation-flow}), i.e., for each \ac{BPMN} sub-process $F'_i$ joining in $G_J$, the method corresponding to $Last(F'_i)$ calls the method corresponding to $G_J$. On the other hand, the method corresponding to $G_J$ calls the method corresponding to the \ac{BPMN} element reached in one step by $G_J$ (i.e., $First(F_J)$, see the last three methods in Figure~\ref{generic_gateway.bpmn.java}).

\begin{figure}
\centerline{\includegraphics[width=0.9\hsize]{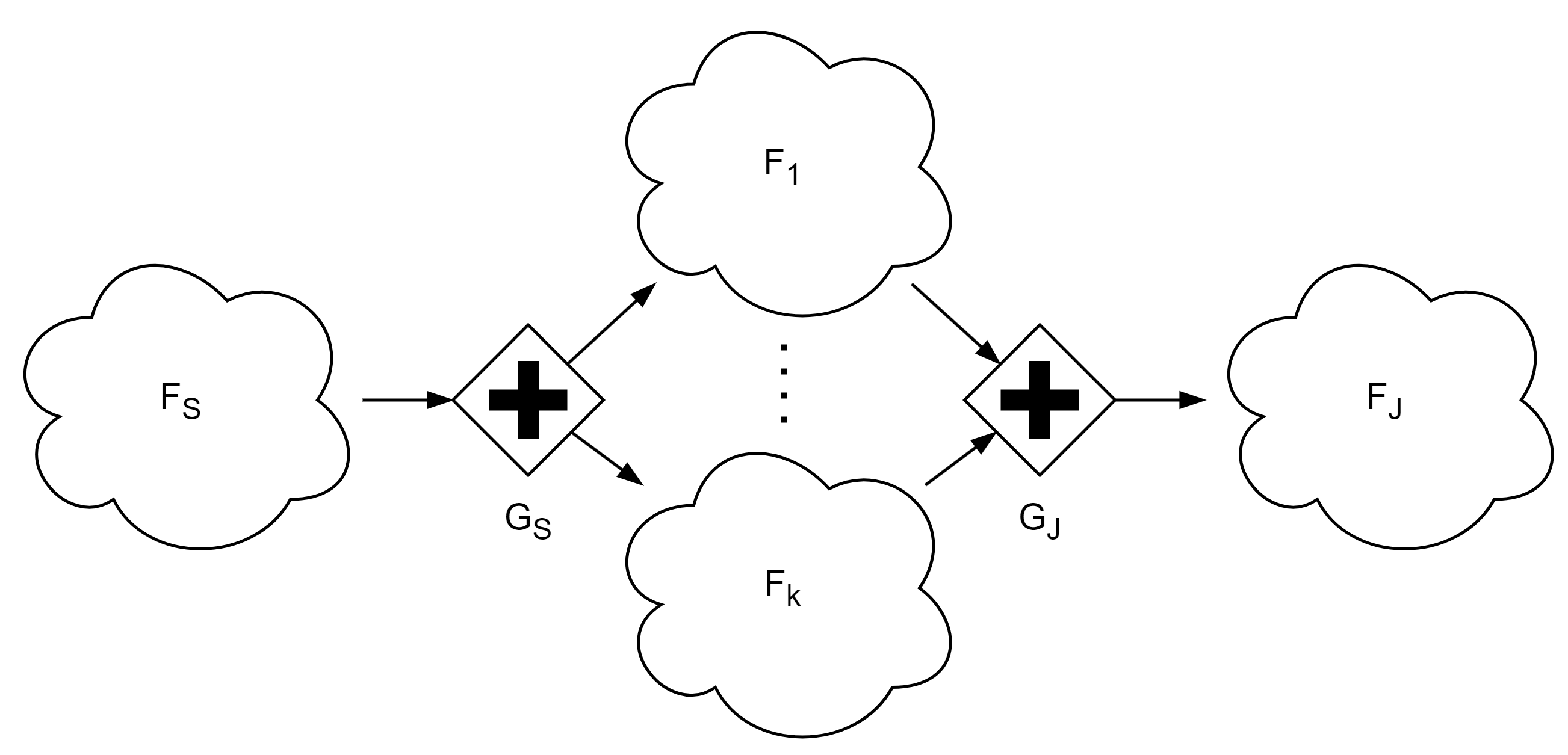}}
\vspace*{8pt}
\caption{A generic \ac{BPMN} Process \Example{GenericPAR} with parallel gateways.
\label{generic_parallel.bpmn}
}
\end{figure}

\subsubsection{Translation of Parallel and Inclusive Gateways} \label{sec:tojava-generation-gateways-par-incl}

Due to their behaviour, parallel and inclusive gateways are treated differently. 
Namely, Figure~\ref{generic_parallel.bpmn} shows a generic \ac{BPMN} subprocess involving parallel gateways, and Figure~\ref{generic_parallel.bpmn.java} shows the corresponding translation.

\begin{figure}
\begin{lstlisting}[language=Java,frame=tlbr]
 class GenericPAR { 
  $\ldots$
 public void TASK_$Last(F_S)$(BeL.ProcessStatus s) {
  GATEWAY_$G_S$(s.withCurrent($Last(F_S)$));
 }
 public void GATEWAY_$G_S$(BeL.ProcessStatus s) {
  BeL.fork(s,"$G_S$",this::TASK_$First(F_1)$,$\ldots$,this::TASK_$First(F_k)$);
  BeL.stopThread();
 }
 $\ldots$
 public void TASK_$Last(F_1)$(BeL.ProcessStatus s) {
  GATEWAY_$G_J$(s.withCurrent($Last(F_1)$));
 }
 $\ldots$
 public void TASK_$Last(F_k)$(BeL.ProcessStatus s) {
  GATEWAY_$G_J$(s.withCurrent($Last(F_k)$));
 } 
 public void GATEWAY_$G_J$(BeL.ProcessStatus s) {
  BeL.join(s,"$G_J$", this::TASK_$First(F_J)$);
 }
 $\ldots$
}

\end{lstlisting}
\vspace*{8pt}
\caption{Java code generated for the \ac{BPMN} Process in Figure \ref{generic_parallel.bpmn}.}\label{generic_parallel.bpmn.java}
\end{figure} 


The translation of the parallel splitting gateway $G_S$ consists in calling the \texttt{fork} \BeL  function, which takes as arguments the method names corresponding to the tasks starting each of the parallel branches to be executed ($First(F_1), \ldots, First(F_k)$). As a result of a call to \texttt{fork}, the current thread (i.e., the one leading to $G_S$) is terminated, and $k$ new threads are started, each corresponding to $First(F_1), \ldots, First(F_k)$.

On the other end of the parallel, the parallel joining gateway $G_J$ uses the \texttt{join} \BeL  function to execute (on a new thread) the next task ($First(F_J)$) once all the parallel branches reached it.

\BeL implements \texttt{fork} and \texttt{join} in two ways. One relies on standard Java multithreading (\textit{true parallel}). The other one approximates the parallel by running the parallel branches sequentially. This is useful to easier check errors in the process definition, since it generates sequential traces that are easier to debug, but may lead to deadlocks if synchronization is required (e.g., messages are exchanged between parallel branches). The user may enable sequential execution when needed, whilst true parallel execution is the default.

Finally, inclusive gateways are similar to parallel gateways, but only edges having a true condition are selected for parallel execution.



\subsection{DMN Model Translation} \label{dmn_trans:subsec}\label{sec:tojava-generation-dmn}

In this section, we describe the behaviour of function \FunctionDef{translateDMN} that is called in line~\ref{dmn_start.step} of Algorithm~\ref{main_algo}. We recall that each \ac{DMN} table $D$ contained in the input tables set ${\cal D}$ has the form shown in Figure~\ref{generic.dmn}. 

\begin{figure}
\begin{lstlisting}[language=Java,frame=tlbr]
// wrapper class for the input of DMN table GenTabDT
class dmn_dtable_GenTabDT_arguments {
 public Object $in_1$;
 $\ldots$
 public Object $in_k$;
}
// wrapper class for the output of DMN table GenTabDT
class dmn_dtable_GenTabDT_result {
 Object $out_1$;
 $\ldots$
 Object $out_n$;
 public dmn_dtable_GenTabDT_result(
   Object $out_1$,$\ldots$,Object $out_n$) {
  this.$out_1$ = $out_1$;
  $\ldots$
  this.$out_n$ = $out_n$;
 }
}
// decision code for DMN table GenTabDT
class dmn_dtable_GenTabDT {
 public static dmn_dtable_GenTabDT_result execute(dmn_dtable_GenTabDT_arguments args) {
  Object $in_1$ = args.$in_1$;
  $\ldots$
  Object $in_k$ = args.$in_k$;
  if ($\hat{c}_{1,1}$ && $\ldots$ && $\hat{c}_{1,k}$) {
   return new dmn_dtable_GenTabDT_result($v_{1,1}$,$\ldots$,$v_{1,n}$);
  } else if
    $\ldots$  
  } else if ($\hat{c}_{m,1}$ && $\ldots$ && $\hat{c}_{m,k}$) {
   return new dmn_dtable_GenTabDT_result($v_{m,1}$,$\ldots$,$v_{m,n}$);
  } else {
   return new dmn_dtable_GenTabDT_result($v_{m+1,1}$,$\ldots$,$v_{m+1,n}$);
  }
 }
}
\end{lstlisting}
\vspace*{8pt}
\caption{Java code generated for the \Example{GenTabDT} DMN in Figure \ref{generic.dmn}.}\label{generic.dmn.java}
\end{figure} 

Given this, each decision table $D$ is translated into a set of three classes, as shown in Figure~\ref{generic.dmn.java}: a class encapsulating the table inputs (decision variables, \texttt{dmn\_dtable\_GenTabDT\_arguments}), another for the outputs (derived values, \texttt{dmn\_dtable\_GenTabDT\_result}) and a main class containing the decision logic, \texttt{dmn\_dtable\_GenTabDT}, with a static \texttt{execute} method that is used to execute the decision process. This is performed by $m$ nested {\tt if}s, where each condition $\hat{c}_{i,j}$ represent the Java version of the corresponding \ac{FEEL} condition $c_{i,j}(in_j)$, obtained via \FEELJava.

It is worth noting that the \textit{default case} of a decision table, if present, such as row $m+1$ in Figure \ref{generic.dmn}, becomes the final \texttt{else} statement in the decision code of Figure \ref{generic.dmn.java}. Otherwise, the translator automatically adds a last \texttt{else} statement calling an internal method that ends the decision process with a failure if no other condition matches the given arguments.

\subsection{Translation of Business Rule Tasks} \label{sec:tojava-generation-business}

As already recalled, the bridge between a \ac{BPMN} process ${\cal B}$ and its linked \ac{DMN} tables ${\cal D}$ is provided by business rule tasks, where \ac{DMN} decision tables are invoked in order to assign values to some process variables. Such values must be determined by ``executing'' the \ac{DMN} table, which may depend also on other variables. The assigned variables may then be used, e.g., in a gateway to redirect the execution flow.

\begin{figure}
\centerline{\includegraphics[width=0.6\hsize]{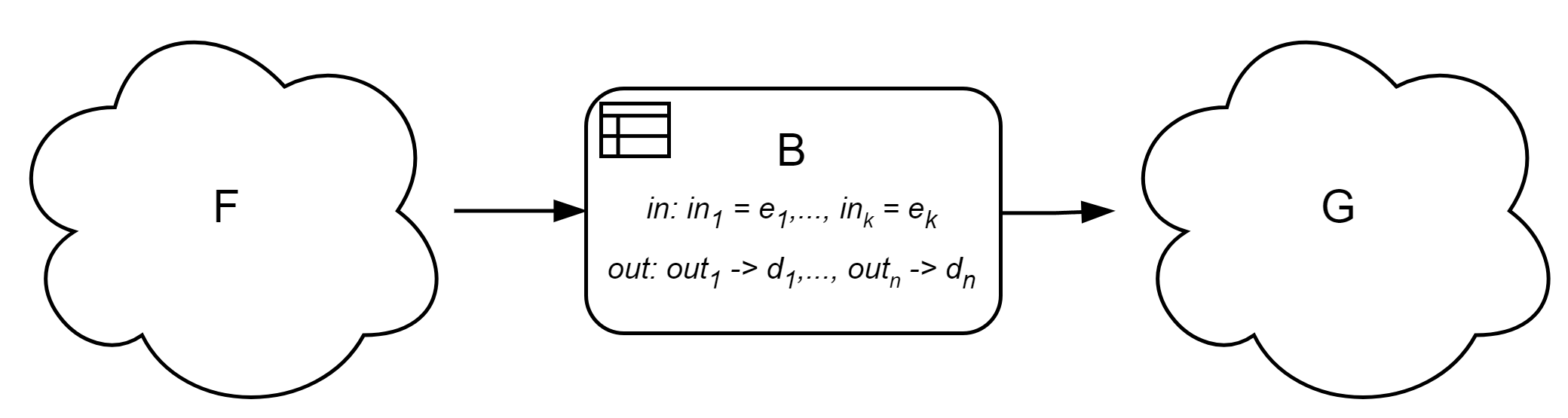}}
\vspace*{8pt}
\caption{A generic \ac{BPMN} Process \Example{GenericBRT} with a business rule task.\label{generic_business.bpmn}}
\end{figure}

Figure~\ref{generic_business.bpmn} shows a generic \ac{BPMN} containing a business rule task. Such task is configured
to invoke the \texttt{GenTabDT} shown in Figure~\ref{generic.dmn}, passing the result of the \ac{FEEL} expressions $e_1,\ldots,e_k$ as the table inputs $in_1,\ldots,in_k$, respectively, and assigning the result values $out_1,\ldots,out_n$ to the process variables $d_1,\ldots,d_k$, respectively.

\begin{figure}
\begin{lstlisting}[language=Java,frame=tlbr]
class GenericBRT {
 $\ldots$
 //Process Variables
 Object $d_1$ = null;
 $\ldots$
 Object $d_n$ = null;
 $\ldots$
 public void TASK_$Last(F)$(BeL.ProcessStatus s) {
  TASK_B(s.withCurrent($Last(F)$));
 } 
 public void TASK_B(BeL.ProcessStatus s) {
   dmn_dtable_GenericTable_arguments args =
     new dmn_dtable_GenericTable_arguments();
   args.$in_1$ = $\hat{e}_1$;
   $\ldots$
   args.$in_k$ = $\hat{e}_k$;
   dmn_dtable_GenericTable_result GenericTableResult =
     dmn_dtable_GenericTable.execute(args);
   $d_1$ = GenericTableResult.$out_1$;
   $\ldots$
   $d_n$ = GenericTableResult.$out_n$;
   TASK_$First(G)$(s.withCurrent("B"));
 }
 $\ldots$
}

\end{lstlisting}
\vspace*{8pt}
\caption{Java code generated for the \ac{BPMN} Process in Figure \ref{generic_business.bpmn}.}\label{generic_business.bpmn.java}
\end{figure} 

This diagram leads to the generation of the code shown in Figure~\ref{generic_business.bpmn.java}, which first packages the table parameters in the table arguments class (using \FEELJava to translate the input expressions $e_i$ to the Java counterpart $\hat{e}_i$), then calls the decision table \texttt{execute} method, and finally unpacks the decision result assigning it to the output process variables $d_1,\ldots,d_k$.



\subsection{Translation of Send and Receive Tasks} \label{sec:tojava-generation-messages}

\begin{figure}
\centerline{\includegraphics[width=0.9\hsize]{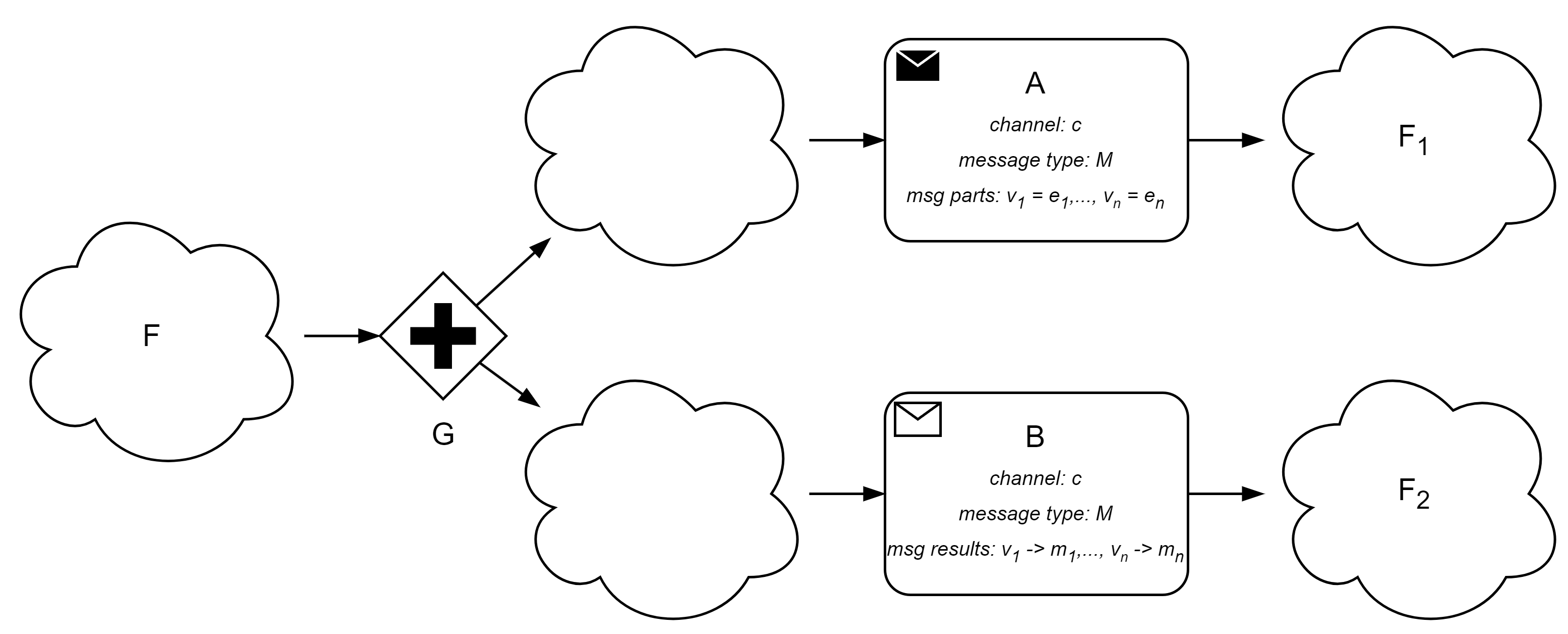}}
\vspace*{8pt}
\caption{A generic \ac{BPMN} Process \Example{GenericMSG} with message exchange through Send and Receive tasks.
\label{generic_message.bpmn}
}
\end{figure}

\BDTrans also supports sending and receiving messages through \textit{send and receive tasks}, which are typically used when two or more \ac{BPMN} branches are executing in parallel.

\begin{figure}
\begin{lstlisting}[language=Java,frame=tlbr]
class GenericMSG { 
 //Process Variables
 Object $m_1$ = null;
 $\ldots$
 Object $m_n$ = null;
 $\ldots$
 //Messages
 private static class Message_$M$ implements BeL.Message {
  Object $v_1$ = null;
  $\ldots$
  Object $v_n$ = null;
 };
 $\ldots$
 public void TASK_A(BeL.ProcessStatus s) {
  Message_$M$ m = new Message_$M$();
  m.$v_1$ = $\hat{e}_1$;
  $\ldots$
  m.$v_n$ = $\hat{e}_n$;
  BeL.sendMessage(s,$c$,m);
  $First(F_1)$(s.withCurrent("A"));
 }
 public void TASK_B(BeL.ProcessStatus s) {
  Message_$M$ m = (Message_$M$)  BeL.receiveMessage(s,$c$);
  $m_1$ = m.$v_1$;
  $\ldots$
  $m_n$ = m.$v_n$;       
  $First(F_2)$(s.withCurrent("B"));
 }
 $\ldots$
}
\end{lstlisting}
\vspace*{8pt}
\caption{Java code generated for the \ac{BPMN} Process in Figure \ref{generic_message.bpmn}.}\label{generic_message.bpmn.java}
\end{figure}

Figure~\ref{generic_message.bpmn} shows a generic Send task \textit{A} and Receive task \textit{B}. Both tasks are configured (in the \ac{BPMN} editor) to use channel $c$ to exchange a message of type $M$, containing parts  $v_1,\ldots,v_n$. As a result, \textit{A} creates an instance of $M$ with $v_i=e_i, i=1 \ldots n$, where $e_i$ are \ac{FEEL} expressions, whereas \textit{B} uses the values $v_1,\ldots,v_n$ from the received instance of $M$ to assign the process variables $m_i,\ldots,m_n$, as described by the Camunda extension elements  \texttt{calledDecision} and \texttt{ioMapping} in the XML serialization of the task. 

To implement the message exchange, the \BeL framework creates a \textit{concurrent blocking queue} for channel $c$ and a class \texttt{Message\_\emph{M}} that declares $v_1,\ldots,v_n$ as member variables, as shown in the generated code in Figure \ref{generic_message.bpmn.java}.

The Send task \textit{A} builds an instance of \texttt{Message\_\emph{M}}, assigns the expressions $\hat{e}_1,\ldots,\hat{e}_n$ (resulting from applying \FEELJava on $e_1,\ldots,e_n$) to its members $v_1,\ldots,v_n$ , and finally
uses the \texttt{sendMessage}  \BeL function to send it on channel $c$.
On the other hand, the Receive task \textit{B} uses the \texttt{receiveMessage} \BeL function to retrieve (possibly blocking if the queue is empty) a message from channel $c$ and then assigns the values from its parts $v_1,\ldots,v_n$ to the process variables $m_1,\ldots,m_n$, respectively.

\subsection{An Example of BPMN Translation} \label{sec:tojava-example}

To illustrate the code generation technique applied to concrete \ac{BPMN}+\ac{DMN} artefacts, we use the \Example{Shipment} as a running example (see Section~\ref{bpmn_descr:subsec} and Figure~\ref{shipment.bpmn}).


\begin{figure}
\begin{lstlisting}[language=Java,frame=tlbr,linerange={105-120}]
 //Input Variables
 // READ: |DMN|GetLengthDT|Type, Activity_0h04jo2
 Object pType = null;
 // READ: |DMN|DetermineModeDT|Weight, Gateway_1tgxmu2, Activity_1cbdv9z, |DMN|ChooseConsentDT|Weight, Activity_1ol43bw
 Object pWeight = null;

 //Process Variables
 // READ: Gateway_0u50uj6
 // WRITTEN: Activity_1cbdv9z
 Object consent = null;
 // READ: |DMN|DetermineModeDT|Length, Activity_1ol43bw, Gateway_0i2yujj
 // WRITTEN: Activity_0h04jo2
 Object pLength = null;
 // READ: |DMN|ChooseConsentDT|Mode, Gateway_1ocbjca, Activity_1cbdv9z
 // WRITTEN: Activity_1ol43bw
 Object sMode = null;
\end{lstlisting}
\vspace*{8pt}
\caption{Generated Variable Declarations for the \Example{Shipment} Process.\label{code_variables.java}}
\end{figure}

\begin{figure}
\centering\framebox{\begin{minipage}{0.8\hsize}
\small
\verbatiminput{shipment/shipment_inputs.txt}
\end{minipage}}
\vspace*{8pt}
\caption{Generated input variables definition for the \Example{Shipment} process.\label{shipment_inputs}}
\end{figure}


The generated running example class contains the process and input variable definitions shown in Figure \ref{code_variables.java}. The input variables are also listed in the generated \textit{inputs file} shown in Figure \ref{shipment_inputs}. 

\begin{figure}
\begin{lstlisting}[language=Java,frame=tlbr,linerange={70-97}]
// wrapper class for the output of DMN table ChooseConsentDT
class dmn_dtable_ChooseConsentDT_result {
 String Consent;
 public dmn_dtable_ChooseConsentDT_result(String Consent) {
  this.Consent = Consent;
 }
}
// wrapper class for the input of DMN table ChooseConsentDT
class dmn_dtable_ChooseConsentDT_arguments {
 public Object Mode;
 public Object Weight;
}
// decision code for DMN table ChooseConsentDT
class dmn_dtable_ChooseConsentDT {
 public static dmn_dtable_ChooseConsentDT_result
     execute(dmn_dtable_ChooseConsentDT_arguments args) {
  Object Mode = args.Mode;
  Object Weight = args.Weight;
  if (BeL.tostring(Mode).equals("car") &&
      BeL.tonumber(Weight) > BeL.tonumber(6.0)) {
   return new dmn_dtable_ChooseConsentDT_result("owner");
  } else if (BeL.tostring(Mode).equals("truck") &&
             BeL.tonumber(Weight) > BeL.tonumber(8.0)) {
   return new dmn_dtable_ChooseConsentDT_result("com");
  } else {
   return new dmn_dtable_ChooseConsentDT_result("none");
  }
 }
\end{lstlisting}
\vspace*{8pt}
\caption{Java Code for the \texttt{ChooseConsentDT} Decision Table (see Figure~\ref{choose_consent.dmn}).\label{choose_consent.java}}
\end{figure}

Figure~\ref{choose_consent.java} shows an example of decision table translation, corresponding to the \texttt{ChooseConsentDT} decision table illustrated in Figure~\ref{choose_consent.dmn}.

\begin{figure}
\begin{lstlisting}[language=Java,frame=tlbr,linerange={145-151}]
 public void EVENT_StartEvent_1_package_received(
             BeL.ProcessStatus s) {
  //Start Event package received [StartEvent_1]
  //[outgoing edge] Activity_0h04jo2 - get length
  TASK_Activity_0h04jo2_get_length(
    s.withCurrent("StartEvent1"));
 }
\end{lstlisting}
\vspace*{8pt}
\caption{Generated Java Code: Start Event.\label{code_start_event.java}}
\end{figure}

The \ac{BPMN} start Event \texttt{StartEvent\_1} is encoded in method\\ \texttt{EVENT\_StartEvent\_1\_package\_received} shown in Figure \ref{code_start_event.java}. Note that the reported code has been simplified by omitting the internal calls to utility methods that generate and display the execution trace, as we will see in Section \ref{sec:execution-trace}.

\begin{figure}
\begin{lstlisting}[language=Java,frame=tlbr,linerange={168-184}]
 public void GATEWAY_Gateway_0u50uj6(BeL.ProcessStatus s) {
  //Exclusive Gateway Gateway_0u50uj6
  if (consent.equals("com")) {
   //[outgoing edge] Activity_1njskid - sign declaration
   TASK_Activity_1njskid_sign_declaration(
     s.withCurrent("Gateway_0u50uj6"));
  } else if (consent.equals("owner")) {
   //[outgoing edge] Activity_1nfni4r - fetch declaration
   TASK_Activity_1nfni4r_fetch_declaration(
     s.withCurrent("Gateway_0u50uj6"));
  } else if (consent.equals("none")) {
   //[outgoing edge] Gateway_07f90ke
   GATEWAY_Gateway_07f90ke(s.withCurrent("Gateway_0u50uj6"));
  } else {
   BeL.noDefaultCaseError(s);
  }
 }
\end{lstlisting}
\vspace*{8pt}
\caption{Generated Java Code: Exclusive Gateway.\label{code_splitting_gateway.java}}
\end{figure}

\begin{figure}
\begin{lstlisting}[language=Java,frame=tlbr,linerange={153-158}]
 public void GATEWAY_Gateway_07f90ke(BeL.ProcessStatus s) {
  //Exclusive Joining Gateway Gateway_07f90ke
  //[outgoing edge] Event_1pjc4df - ready for shipment
  EVENT_Event_1pjc4df_ready_for_shipment(
     s.withCurrent("Gateway_07f90ke"));
 }
\end{lstlisting}
\vspace*{8pt}
\caption{Generated Java Code: Joining gateway.\label{code_joining_gateway.java}}
\end{figure}

Gateways such as the splitting exclusive after the \texttt{choose consent} task and the joining exclusive before the end event in Figure~\ref{shipment.bpmn} are rendered as shown in Figures \ref{code_splitting_gateway.java} and \ref{code_joining_gateway.java}, respectively, following the already discussed rationale.

\begin{figure}
\begin{lstlisting}[language=Java,frame=tlbr,linerange={216-228}]
 public void TASK_Activity_1cbdv9z_choose_consent(
        BeL.ProcessStatus s) {
  //Business Rule Task choose consent [Activity_1cbdv9z]
  dmn_dtable_ChooseConsentDT_arguments args =
    new dmn_dtable_ChooseConsentDT_arguments();
  args.Mode = sMode;
  args.Weight = pWeight;
  dmn_dtable_ChooseConsentDT_result chooseConsentResult =
    dmn_dtable_ChooseConsentDT.execute(args);
  consent = chooseConsentResult.Consent;
  //[outgoing edge] Gateway_0u50uj6
  GATEWAY_Gateway_0u50uj6(s.withCurrent("Activity_1cbdv9z"));
 }
\end{lstlisting}
\vspace*{8pt}
\caption{Generated Java Code:  Business Rule Task using the \texttt{ChooseContentDT} Decision Table.\label{code_business_task.java}}
\end{figure}

Figure \ref{choose_consent.java} shows the code generated by the Business Rule task \texttt{choose consent}.



\begin{figure}
\begin{lstlisting}[language=Java,frame=tlbr,linerange={133-143}]
 public void EVENT_Event_19ylwnc_no_shipment(
        BeL.ProcessStatus s) {
  //End Event no shipment [Event_19ylwnc]
  BeL.error(s, "No Shipment", 3);
 }

 public void EVENT_Event_1pjc4df_ready_for_shipment(
        BeL.ProcessStatus s) {
  //End Event ready for shipment [Event_1pjc4df]
  BeL.success(s);
 }
\end{lstlisting}
\vspace*{8pt}
\caption{Generated Java Code: End Events.\label{code_end_event.java}}
\end{figure}

Finally, Figure \ref{code_end_event.java} shows the two end events of the running example: In particular, when an exception end event is activated, the code passes the event description and code (if available) to the \textit{error} \BeL  function, as for the \texttt{EVENT\_Event\_19ylwnc\_no\_shipment}  in Figure \ref{code_end_event.java}, in order to signal the specific exception occurred in the execution trace.






\subsection{Main Implementation Details}\label{implementation:subsec}

In this section, we provide some implementation details which are relevant for the translation part discussed here.
\BDTrans takes as input \ac{BPMN} 2.0 diagrams and \ac{DMN} decision models serialized in their standard XML format \citep{bpmn,dmn}, possibly with some \texttt{extensionElements} introduced by the Camunda Modeler \citep{camundamodeler}, mainly used to map inputs and output variables in various elements.
For sake of simplicity and without loss of generality, we suppose that each \ac{BPMN} file defines a single business process (though \BDTrans is able to handle also multiple processes in a single file), whereas the decision tables referenced by such process can be described in one or more \ac{DMN} files.

The generated output consists of Java classes and support files (used to implement \BeLNoSp) that can be used for documentation, simulation, and verification purposes (the latter part is actually performed by the \BDTest algorithm, see Figure~\ref{overall_process} and Section~\ref{sec:analysis}), as well as a skeleton for ad-hoc process development.

If a valid Java compiler is found in the system, \BDTrans also compiles the generated Java artifacts (which may produce a large number of classes) and packages them in an auto-executable JAR file. For simplicity, the compiled classes are placed in the root package, while the JAR contains two \texttt{src} and \texttt{doc} folders where the algorithm places the code sources and the other generated artifacts, respectively.

\subsubsection{Process Graph and Execution Trace} \label{sec:execution-trace}

\begin{figure}
\centering\framebox{\begin{minipage}{0.8\hsize}
\small
\verbatiminput{shipment/shipment_execution_output_1.txt}
\end{minipage}}
\vspace*{8pt}
\caption{An example of execution output obtained running the \Example{Shipment} process.\label{shipment_output_1}}
\end{figure}

The Java program generated by \BDTrans, if executed, outputs a detailed execution description such as the one reported in Figure~\ref{shipment_output_1}.
The output reports the values assigned to the input variables and lists all the \ac{BPMN} nodes encountered in execution order. When a business rule task is activated, the output also shows the decision table name and its results. Finally, when an end event is reached, the output reports the end status (success in this case).

\begin{figure}
\centering\framebox{\begin{minipage}{0.8\hsize}
\small
\verbatiminput{shipment/shipment_outputs.txt}
\end{minipage}}
\vspace*{8pt}
\caption{Generated output status after running an instance of the \Example{Shipment} process.\label{shipment_outputs}}
\end{figure}

However, while this output is designed to be human-readable, the model execution also generates an execution summary file useful to be automatically parsed. Figure~\ref{shipment_outputs} shows such file after an example run of the \Example{Shipment} process. The file reports both the inputs used and the its output in terms of message, code and success status.



\begin{figure}
\centering\framebox{\begin{minipage}{0.8\hsize}
\small
\verbatiminput{shipment/shipment.trace}
\end{minipage}}
\vspace*{8pt}
\caption{An example of execution trace generated by the \Example{Shipment} process.\label{shipment.trace}}
\end{figure} 

To this aim, the \BeL framework can be configured to generate \textit{trace files} that are actually a simplified version of the execution output shown in Figure \ref{shipment_output_1} and report the nodes and edges of the process graph activated in execution order. The trace format is the same as the graph file, to simplify comparison. An example is shown in Figure \ref{shipment.trace}.

\subsection{Current Limitations}\label{limitations:subsec}\label{subsec:limits}

In this section we describe some current limitations of \BDTransNoSp, which we plan to overcome as future work (see Section~\ref{sec:conclusions}).
We point out that, even with such limitations, our approach is effective as shown by our experimental results (see Section~\ref{sec:expres}).

A first limitation is given by \ac{BPMN} processes in which a generic user task sets an input variable within a \ac{BPMN} loop (e.g., suppose that the ``measure weight'' task in Figure~\ref{shipment.bpmn} goes back to the ``get length'' task). In fact, the current version of \BDTrans will reuse the same value in each iteration, even if it may potentially change. Solving such issue will entail detecting user tasks with such a behaviour, and handling them in a suitable way (i.e., many inputs value have to be provided for the same variable). 

A second limitation impacts on the performance of the generated Java program. Namely,  the current execution model involves nested method calls for each transition, so limiting the overall length of the execution traces to the size of the language call stack. While \acp{BPMN} are typically small-sized, this can become a performance issue for the execution of \ac{BPMN} involving loops to be traversed multiple times. In order to overcome such limitation, a different translation scheme may be used when process execution efficiency is important (e.g., by employing a main execution loop which may handle a generic \ac{BPMN} transition).  

As a third limitation, \BDTrans currently supports parallelism only in the setting shown in Figure~\ref{generic_parallel.bpmn}. Instead, \ac{BPMN} also allows usage of event-based gateways to decide when to branch threads basing on other threads actions (send/receive messages, signals and so on). Furthermore, we only support synchronization via message exchange, while \ac{BPMN} also allows other ways of communication (e.g., signals). In order to overcome such limitation, we need to address the full \ac{BPMN} support to parallelism. 

A fourth limitation regards the \ac{BPMN} elements which we do not consider in this paper, i.e., event-based gateways and inclusive/parallel gateways. In event-based gateways, edge conditions are not a \ac{FEEL} Boolean expressions, but are based on events occurring outside the \ac{BPMN} process. This is not a severe limitation, as event-based gateways may be modeled by introducing a new task setting a new input variable $v$, and then replacing the event-based gateway with an exclusive one, whose edges have conditions on $v$. As for inclusive/parallel gateways, \BDTransTest already supports them (see Section~\ref{sec:tojava-generation-gateways-par-incl}), but we leave an evaluation of \ac{BPMN}+\ac{DMN} processes including parallel parts as future work (see Section~\ref{sec:conclusions}). This also holds for tasks including send and receive operations, which are typically used when dealing to parallel \acp{BPMN}.
Moreover, \BDTrans disregards \ac{BPMN} pools and lanes, if present. In fact, such \ac{BPMN} elements are meant as annotations to recall who is executing a set of tasks, thus it is not important in the verification problem, our focus here. 

However, even with the aforementioned limitations due to its current development phase, \BDTrans is already capable to translate realistic and complex BPMN and DMN artefacts, as shown by the examples reported in this paper.

\subsection{\BDTrans Correctness}\label{proof:subsec}

To conclude this section, we discuss 
the correctness for our translation procedure \BDTrans (Algorithm~\ref{main_algo}), assuming the input \ac{BPMN} does not use parallel nor inclusive gateways.


By borrowing ideas both from model checking~\cite{mc_book} and the notion of bisimulation~\cite{San2011}, let $S = T \times {\cal A}$ be the {\em state space} of the input \ac{BPMN}+\ac{DMN} $({\cal B}, {\cal D})$, where $T$ is the set of nodes (tasks, events and gateways) of ${\cal B}$ and ${\cal A}$ the set of all values for each variable of ${\cal B}$, including the special undefined value $\perp$. Thus, if $v_1, \ldots, v_k$ are the variables of $({\cal B}, {\cal D})$ with domains $D_1, \ldots, D_k$, then ${\cal A} = (D_1 \cup \{\perp\}) \times \ldots \times (D_k \cup \{\perp\})$. We also define a function $\sigma$ to extract the \ac{BPMN} state from a transition system state, and a function $\mu$ to extract the variables values part. As an example, for the \ac{BPMN} in Figure~\ref{shipment.bpmn}, a possible state is
$s = \langle {\rm get\ length}, \langle{\rm pType} \gets {\rm xl}, {\rm pWeight} \gets 9.5, {\rm pLength} \gets {\perp,} {\rm consent} \gets \perp, {\rm sMode} \gets \perp\rangle\rangle$,
and $\sigma(s) = $get length. This allows us to the define the {\em transition relation} of $({\cal B}, {\cal D})$ as $\delta$, such that $(s, t) \in \delta$ iff i) there is an edge $w = (\sigma(s), \sigma(t))$ in ${\cal B}$, ii) $\mu(s)$ satisfies the condition on $w$, if any, and iii) $\mu(t)$ is obtained by modifying $\mu(s)$ accordingly to actions to be taken on $s$ (possibly depending on tables in ${\cal D}$). In the same example above, if we name the gateway between the ``get length'' and ``measure weight'' tasks as $G_1$, we have that $(s, t), (t, u) \in \delta$, being
$t = \langle G_1, \langle{\rm pType} \gets {\rm xl}, {\rm pWeight} \gets 9.5, {\rm pLength} \gets 2, {\rm consent} \gets \perp, {\rm sMode} \gets \perp\rangle\rangle$ and $\sigma(u) = {\rm measure\ weight}, \mu(u) = \mu(t)$.
This also defines {\em paths} of $({\cal B}, {\cal D})$ as sequences of states $s_1, \ldots, s_k$ such that $(s_i, t_i) \in \delta$ for all $i = 1, \ldots, k$, being $\sigma(s_1)$ a start event of ${\cal B}$. 
Note that, in our definition, a path may not terminate in an end event. 

On the other hand, we define the {\em trace} of the Java class $J({\cal B}, {\cal D})$ as the output of $J({\cal B}, {\cal D})$. 
More formally, such a trace consists of a sequence $\langle (n_1, \nu_1)\ldots, \ldots, (n_k, \nu_k)\rangle$, where $n_i$ are nodes of ${\cal B}$ and $\nu_i$ contains one value (possibly $\perp$) for all variables of $({\cal B}, {\cal D})$. 

Theorem~\ref{main_theorem} states the correctness of our translation; its proof follows from how the translation itself is carried out.

\begin{theorem}\label{main_theorem}
  Let $({\cal B}, {\cal D})$ be a \ac{BPMN} and a set of \ac{DMN} tables linked to it, respectively, with variables $v_1, \ldots, v_k$. Suppose ${\cal B}$ fulfills the limitations of Section~\ref{limitations:subsec} and does not use neither parallel nor inclusive gateways, and
  let $J({\cal B}, {\cal D})$ be the result of \FunctionDef{BDTrans} in Algorithm~\ref{main_algo} when applied to $({\cal B}, {\cal D})$.
  Then, for each path $s_1, \ldots, s_k$ of $({\cal B}, {\cal D})$, there exists one unique trace $\langle (n_1, \nu_1)\ldots, \ldots, (n_k, \nu_k)\rangle$ of  $J({\cal B}, {\cal D})$ s.t. $\sigma(s_i) = n_i$ and $\mu(s_i) = \nu_i$ for all $i = 1, \ldots, k$. The viceversa also holds.
\end{theorem}

%% file: post_processing.tex
\section{\ac{BPMN}+\ac{DMN} Testing and Coverage Analysis} \label{sec:analysis}

In this section, we describe the second part of our \BDTransTest tool, i.e., the algorithm which takes as input a set of Java classes describing a given BPMN+DMN $({\cal B}, {\cal D})$, a description of the domains for the input variables, and performs thorough testing of  $({\cal B}, {\cal D})$, also measuring the achieved coverage. From now on, we will refer to such algorithm as \BDTestNoSp, as depicted in Figure~\ref{overall_process}.

The rest of this section is organized as follows. Section~\ref{sec:tojava-inputs} describes the second input of \BDTestNoSp, i.e., the information on the domains of the input variables. Section~\ref{bpmn_testing:subsec} describes how the verification is performed.

\subsection{BPMN Inputs Analysis} \label{sec:tojava-inputs}

As already anticipated in Section~\ref{overall:subsec} and Figure~\ref{overall_process}, \BDTest also needs a description of the input variables used in $({\cal B}, {\cal D})$, together with all possible information on their domains. This is accomplished by function \textsl{getVarsDoms}, called in line~\ref{get_vars.step} of Algorithm~\ref{main_algo}, which returns a set $V_I$ of triplets $(v, t, d)$, where $v$ is a variable, $t$ its Java type and $d$ contains the information that is actually needed at this stage.
Namely, $d$ may be of 4 different types:
\begin{itemize}
\item \textsl{ENUM}$(E)$, being $E$ a finite set of constant values, which means that variable $v$ is an enumerative variable which may only take values in $E$. This is caused by $v$ being only used in direct equality comparisons within the input \ac{BPMN}.
\item \textsl{BALL}$(w)$, being $w$ a numeric value, which means that variable $v$ takes values in a ball centered in $w$. This is caused by $v$ being used in gateways with \ac{FEEL} expression {\tt $v$ <= $w$} and its corresponding ``default'' case.
\item \textsl{RANGE}$(I)$, being $I$ a numeric range (both ends may be either included or excluded), which means that variable $v$ takes values in the range $I$. This is caused by $v$ being used in gateways with \ac{FEEL} expression {\tt $v$ in $I$} or in \ac{DMN} employing $I$.
\item \textsl{UNHANDLED}$(F)$, being $F$ a \ac{FEEL} expression, which means that $F$ cannot be automatically understood by \textsl{getVarsDoms}, i.e., if $F$ does not fall into one of the above cases. \BDTest allows the developer to provide the correct domain for such variables.
\end{itemize}


\begin{figure}
\centering
\framebox{\begin{minipage}{0.28\hsize}        
    $v>11$, $v<=50$
    
    $p="yes"$, $p="no"$
    
    $w>0$, $w<abs(v)$
\end{minipage}}\quad%
\framebox{\begin{minipage}{0.62\hsize}
\small
\verbatiminput{other_code/generic_inputs.txt}
\end{minipage}}
\vspace*{8pt}
\caption{An example of FEEL expressions on input variables (left) and the corresponding generated inputs file (right).\label{generic.inputs}}
\end{figure}

The information in set $V_I$ output by \textsl{getVarsDoms} is then written to an external \textit{inputs file}. For each variable, such file also contains a random value, compatible with the determined domain. This file will be read by the model code during the initialization phase, in order to assign a value to all its input variables.

As an example, Figure \ref{generic.inputs} shows some FEEL expressions involving input variables $v,p,w$ and the corresponding generated inputs file.

\subsection{Overall \ac{BPMN} Testing} \label{bpmn_testing:subsec}

Our post-process algorithm \BDTest takes as inputs the JAR file $J({\cal B}, {\cal D})$, the input variables file $V({\cal B}, {\cal D})$ and the underlying graph file $G({\cal B})$ which are output by \BDTransNoSp. 
As a result, \BDTest outputs {\em FAIL} or {\em PASS}, together with the achieved coverage percentages $C_n \in [0, 100]$ on ${\cal B}$ blocks and $C_e \in [0, 100]$ on ${\cal B}$ edges. 

To this aim, \BDTest is based on a subroutine {\sl RunOnce}. Namely, {\sl RunOnce} parses the input variables file $V({\cal B}, {\cal D})$, and assigns to each input variable in $V({\cal B}, {\cal D})$ a random value taken from the corresponding domain (that can be again found in $V({\cal B}, {\cal D})$). Then, it uses such randomly-generated values to execute $J({\cal B}, {\cal D})$. Since such an execution may turn out to be non-terminating (e.g., because of a loop in ${\cal B}$), a configurable timeout $\tau$ is set for the execution itself. Finally, once {\sl RunOnce} is terminated, \BDTest uses its outputs and the graph file $G({\cal B})$ to keep track of which nodes and which edges have been traversed at least once. This is used to provide the final nodes coverage and branch coverage w.r.t. ${\cal B}$.

\BDTest has several invocation modes, which are turned on by run-time options. In the following, we describe the most important ones.

\begin{itemize}
\item Repeatedly invoke {\sl RunOnce} for $k \leq n$ times, for a given $n \in \mathbb{N}$, stopping before $n$ executions are made (i.e., with $k < n$) if $C_n \geq \theta_n \land C_e \geq \theta_e$ for given thresholds $\theta_n, \theta_e \in [0, 100]$. This 
  allows users
  to avoid ``useless'' executions once a given desired coverage has been reached. \BDTest also implements the following variants of such modality, in order to provide more flexibility to users: i) only $C_n \geq \theta_n$ is checked; ii) only $C_e \geq \theta_e$ is checked; iii) $C_n \geq \theta_n \lor C_e \geq \theta_e$ is checked. The final result is {\em PASS} iff the selected condition on coverages is fulfilled at the end of the $k$ {\sl RunOnce} invocations, and {\em FAIL} otherwise. Note that, as usual in testing, if the output is {\em PASS} then there exists $k$ executions yielding the desired coverage, but the viceversa does not hold. The dual property also holds: if there does not exist $k$ executions of ${\cal B}$ achieving the desired coverage, then \BDTest outputs {\em FAIL}, but the viceversa is not true.
  \item Repeatedly invoke {\sl RunOnce} for $k \leq n$ times, for a given $n \in \mathbb{N}$, stopping before $n$ executions are made (i.e., with $k < n$) if $J({\cal B}, {\cal D})$ reports that a BPMN error block is traversed. In this case, as it is usual in testing, if the output is {\em FAIL} then there exists an execution of ${\cal B}$ which traverses an error block of ${\cal B}$, but the viceversa does not hold. The dual property also holds: if there does not exist an execution of ${\cal B}$ which traverses an error block of ${\cal B}$, then \BDTest outputs {\em PASS}, but the viceversa is not true.
  \item Use a statistical model checking algorithm to decide the number of {\sl RunOnce} invocations $n$. In this case, two real numbers $\varepsilon, \delta \in (0, 1)$ must be provided. \BDTest allows the user to choose between two different properties to be verified: i) a BPMN error block is never traversed and ii) it is not possible to achieve overall coverage $C_n \geq \theta_n \land C_e \geq \theta_e$ for given thresholds $\theta_n, \theta_e \in [0, 100]$ (the same 3 variants for the coverage check described  above may be used as well). As a result, the probability that the final output is {\em PASS} when it should be {\em FAIL} is at most $\varepsilon$ with confidence $1 -\delta$. Note that 
    only type-I error are possible, i.e., if the final output is {\em FAIL} then it is always correct, while an output {\em PASS} may be wrong. 
\end{itemize}



%% file: execution.tex
\section{Experimental Evaluation} \label{sec:expres}

\begin{figure}
\centerline{\includegraphics[width=\hsize]{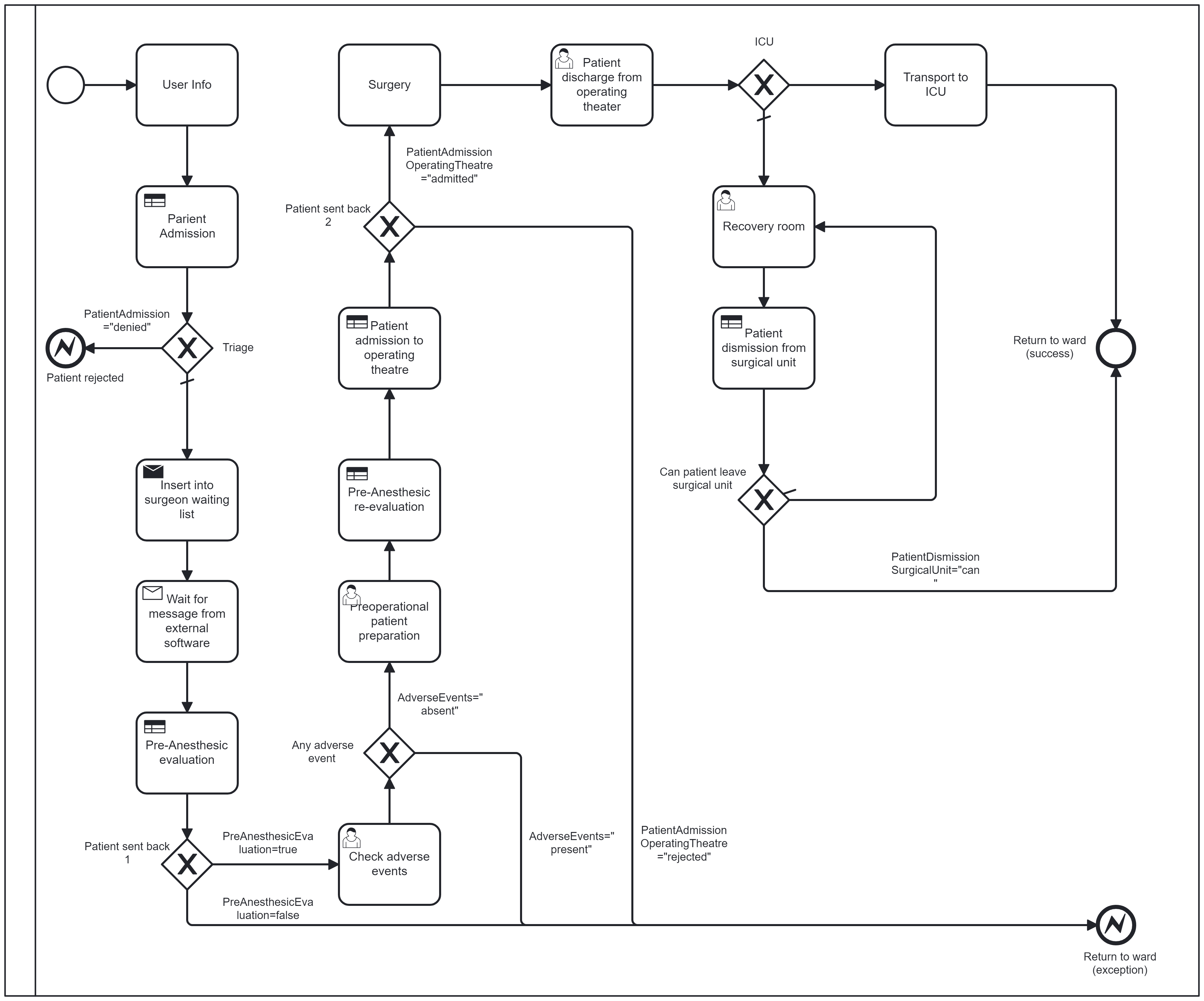}}
\vspace*{8pt}
\caption{\textit{Surgery} Business Process from~\citep{BMPP21}.\label{new_surgery.bpmn}}
\end{figure}

\begin{figure}
\centerline{\includegraphics[width=\hsize]{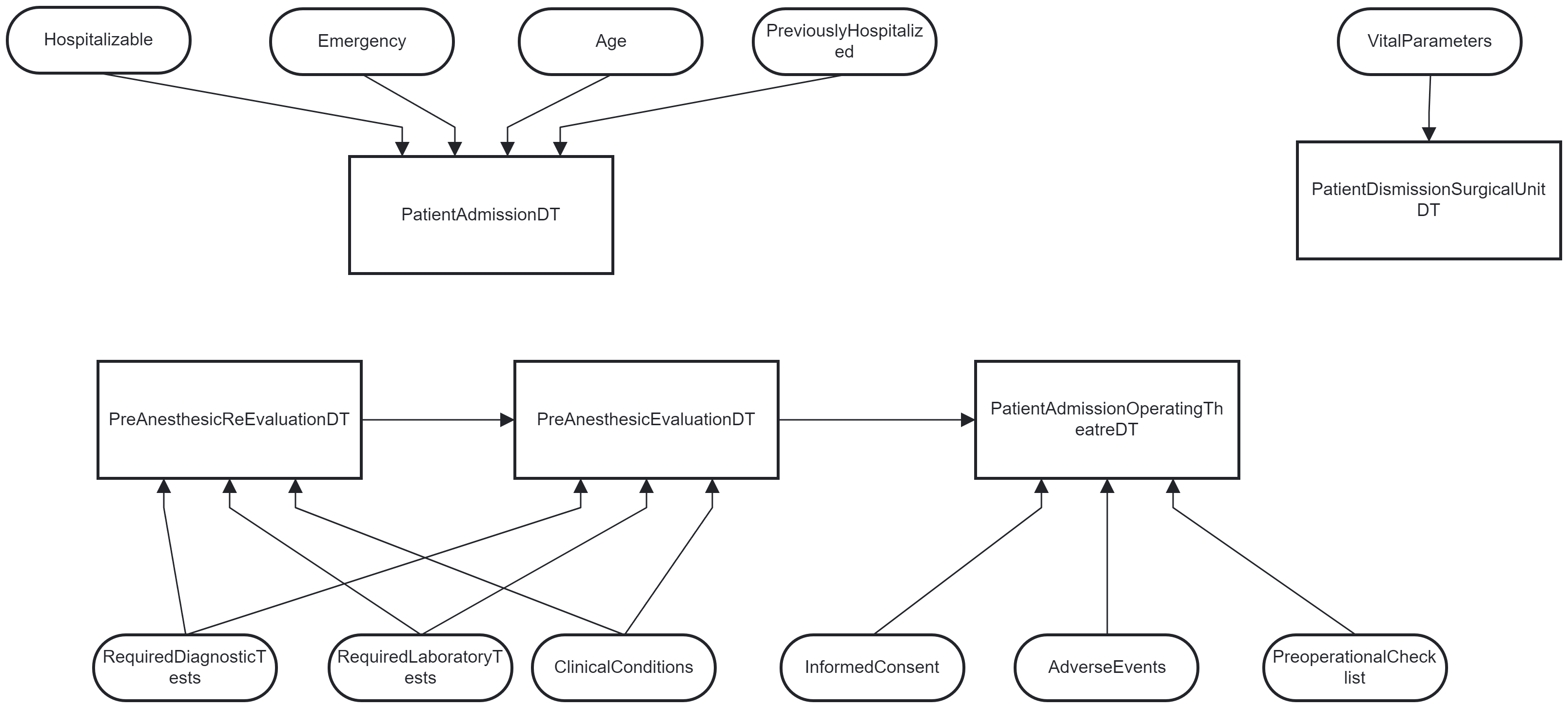}}
\vspace*{8pt}
\caption{Graphical summary of the \textit{Surgery} Decision Model, adapted from~\citep{BMPP21}.\label{new_surgery.dmn}}
\end{figure}

In order to evaluate the effectiveness of our methodology, we run our \BDTransTest tool on two case studies. The first case study is the \Example{Shipment} \ac{BPMN} process described in Section~\ref{bpmn_descr:subsec}; the second case study is the \Example{Surgery} \ac{BPMN}+\ac{DMN} process from~\cite{BMPP21}, which describes the decisions to be taken when operating a patient (see Figures~\ref{new_surgery.bpmn} and~\ref{new_surgery.dmn}).

For both such case studies, the experimental evaluation is organized so as to: i) generate the Java code from the input BPMN+DMN files; ii) run 1000 test cases, using a random numbers generator for the inputs (i.e., we run 1000 times the \FunctionDef{RunOnce} subroutine discussed in Section~\ref{sec:analysis}). All our experiments are run on an Intel i9 with 2.5GhZ and 64GB of RAM.

Table~\ref{expres.tab} summarizes our results. The first column of Table~\ref{expres.tab} shows the resource measurement we consider, i.e.: the time and RAM required for the completion of Java code generation and test cases execution, the time and RAM for Java code generation from BPMN and DMN files only, the average and standard deviation for test cases execution time, the maximum RAM required by all test case executions, and finally the coverage achieved in terms of nodes and edges percentage. The other two columns of Table~\ref{expres.tab} show the corresponding values for the given case study, i.e., \Example{Surgery} for the second column and \Example{Shipment} for the third column.

As a result, times (and RAM) requirements for Java code generation and test cases setting and execution are negligible, while allowing to complete a verification phase with at least 70\% coverage in a matter of a few minutes. This shows the feasibility of our approach.

\begin{table}
  \centering
  \caption{Experimental Results on \Example{Surgery} and \Example{Shipment} case studies}\label{expres.tab}
  \begin{tabular}{|c||c|c|}
    \hline
    {\bf Measure} & {\bf Surgery} & {\bf Shipment}\\\hline\hline
    Total time (mm:ss) & 2:44 & 2:41\\\hline
    Total RAM (MB) & 28.4 & 28.4\\\hline
    Java generation time (secs) & 1.32 & 1.03\\\hline
    Java generation RAM (MB) & 189 & 172\\\hline
    Avg test exec time (secs) & 0.1 & 0.1\\\hline
    Stddev test exec time (secs) & 0.006 & 0.006\\\hline
    Max test exec RAM (MB) & 57.3 & 56.9\\\hline
    Nodes coverage & 87.5\% & 75.0\%\\\hline
    Edges coverage & 81.5\% & 70.6\%\\\hline
\end{tabular}
\end{table}





%% file: conclusion.tex
\section{Conclusions and Future Work}\label{sec:conclusions}

In this paper we introduced {\sf BDTransTest}, a novel methodology and tool to perform execution and verification of business processes described using \ac{BPMN} and also encompassing \ac{DMN} tables. Namely, {\sf BDTransTest} first translates the input \ac{BPMN}+\ac{DMN} files ${\cal B}, {\cal D}$ in an executable Java class, then exploits the generated Java class to perform testing of the given ${\cal B}, {\cal D}$.

We remark that the translation part is valuable also if taken alone, as it allows both static inspection and framework-free execution of ${\cal B}, {\cal D}$. On the contrary, current \ac{BPMN} frameworks like, e.g., Camunda, does not provide details on the code being used to simulate a process, and the simulation itself requires the full framework to be active and running.

As for the verification part, it allows to detect process input variables and guess relevant input values for such variables. Our experimental results shows that, on our case studies, 1000 test cases are enough to obtain more than 70\% coverage in terms of both nodes and edges in the original \ac{BPMN} process. This shows the effectiveness of our approach.


Finally, we outline our main lines of future work on {\sf BDTransTest}: i) overcome all limitations discussed in Section~\ref{subsec:limits}; ii) enrich the \ac{BPMN} verification part by measuring glass-box control-flow-graph-based testing coverages like loop boundary~\citep{WJD13} and definition-use pairs adequacy criteria~\citep{RW85}; iii) enrich the \ac{DMN} verification part by measuring \ac{MCDC} coverage~\citep{Hay01}.




%% file: bpmn.bib
@misc{bpmn,
    author      = "OMG",
    title       = "{{Business Process Model and Notation (BPMN) Version 2.0}}",
    year        = "2011",
    url         = "https://www.omg.org/spec/BPMN/2.0/PDF"
}

@misc{dmn,
    author       = "OMG",
    title        = "{{Decision Model and Notation Version 1.3}}",
    year         = "2019",
    url          = "https://www.omg.org/spec/DMN/1.3/PDF"
}

@misc{feel,
    author       = "OMG",
    title        = "{{Friendly Enough Expression Language}}",
    year         = "2019",
    url          = "https://docs.camunda.org/manual/latest/reference/dmn/feel/"
}

@article{AG16,
title = {Verification of BPMN 2.0 Process Models: An Event Log-based Approach},
journal = {Procedia Computer Science},
volume = {100},
pages = {1064--1070},
year = {2016},
ignorenote = {International Conference on ENTERprise Information Systems/International Conference on Project MANagement/International Conference on Health and Social Care Information Systems and Technologies, CENTERIS/ProjMAN / HCist 2016},
issn = {1877-0509},
doi = {https://doi.org/10.1016/j.procs.2016.09.282},
ignoreurl = {https://www.sciencedirect.com/science/article/pii/S1877050916324516},
author = {Olfa Allani and Sonia Ayachi Ghannouchi},
}

@InProceedings{BGB20,
author="Brumbulli, Mihal
and Gaudin, Emmanuel
and Berre, Fr{\'e}d{\'e}ric",
editor="Boy, Guy Andr{\'e}
and Guegan, Alan
and Krob, Daniel
and Vion, Vincent",
title="Verification of BPMN Models",
booktitle="Complex Systems Design {\&} Management",
year="2020",
publisher="Springer International Publishing",
address="Cham",
pages="27--36",
isbn="978-3-030-34843-4"
}

@InProceedings{BGT20,
author = "Mihal Brumbulli and Emmanuel Gaudin and Ciprian Teodorov",
title = "Automatic Verification of BPMN Models",
booktitle = "10th European Congress on Embedded Real Time Software and Systems (ERTS 2020)",
year = "2020"
}

@InProceedings{KPS17,
author = "Ajay Krishna and Pascal Poizat and Gwen Salaun",
title = "VBPMN: Automated Verification of BPMN Processes",
booktitle = "13th International Conference on integrated Formal Methods (iFM 2017)",
year = "2017"
}

@InProceedings{KDS14,
author = "Faikcan Kog and Attila Dikbas and Raimar J. Scherer",
title = "Verification and validation approach of BPMN represented construction processes",
booktitle = "Creative Construction Conference",
year = "2014"
}

@article{LGT23,
author = "T. Lopes and S. Guerreiro",
year = "2023",
title = "Assessing business process models: a literature review on techniques for BPMN testing and formal verification",
journal = "Business Process Management Journal",
volume = "29",
number = "8",
pages = "133--162",
doi = "https://doi.org/10.1108/BPMJ-11-2022-0557"
}

@article{LFM21,
author = "M. de Leoni and
P. Felli and
M. Montali",
year = "2021",
title = "Integrating BPMN and DMN: Modeling and Analysis",
journal = "Journal on Data Semantics",
pages = "165--188",
volume = "10",
number = "1",
doi = "10.1007/s13740-021-00132-z"
}

@article{intrigila2021lightweight,
  title={A Lightweight BPMN Extension for Business Process-Oriented Requirements Engineering},
  author={Benedetto Intrigila and Giuseppe Della Penna and Andrea D'Ambrogio},
  journal={Computers},
  volume={10},
  number={12},
  pages={171},
  year={2021},
  publisher={MDPI}
}

@Article{IDDCG23,
AUTHOR = {Intrigila, Benedetto and Della Penna, Giuseppe and D’Ambrogio, Andrea and Campagna, Dario and Grigore, Malina},
TITLE = {Process-Oriented Requirements Definition and Analysis of Software Components in Critical Systems},
JOURNAL = {Computers},
VOLUME = {12},
YEAR = {2023},
NUMBER = {9},
ARTICLE-NUMBER = {184},
ignoreURL = {https://www.mdpi.com/2073-431X/12/9/184},
ISSN = {2073-431X},
DOI = {10.3390/computers12090184}
}

@inproceedings{KSRLK24,
  author       = {Tim Kr{\"{a}}uter and
                  Patrick St{\"{u}}nkel and
                  Adrian Rutle and
                  Yngve Lamo and
                  Harald K{\"{o}}nig},
  editor       = {Adela del{-}R{\'{\i}}o{-}Ortega and
                  Marco Montali and
                  Stefanie Rinderle{-}Ma and
                  Hajo A. Reijers and
                  Jan vom Brocke and
                  Mathias Weske and
                  Beno{\^{\i}}t Depaire and
                  Marta Indulska and
                  Han van der Aa and
                  Weronika T. Adrian and
                  Laura Genga and
                  Sander J. J. Leemans and
                  Katarzyna Gdowska and
                  Mar{\'{\i}}a Teresa G{\'{o}}mez{-}L{\'{o}}pez and
                  Jana{-}Rebecca Rehse and
                  Simone Agostinelli},
  title        = {{BPMN} Analyzer 2.0: Instantaneous, Comprehensible, and Fixable Control
                  Flow Analysis for Realistic {BPMN} Models},
  booktitle    = {Proceedings of the Best Dissertation Award, Doctoral Consortium, and
                  Demonstration {\&} Resources Forum at {BPM} 2024 co-located with
                  22nd International Conference on Business Process Management {(BPM}
                  2024), Krakow, Poland, September 1st to 6th, 2024},
  series       = {{CEUR} Workshop Proceedings},
  volume       = {3758},
  pages        = {66--70},
  publisher    = {CEUR-WS.org},
  year         = {2024},
  ignoreurl          = {https://ceur-ws.org/Vol-3758/paper-11.pdf},
  bibsource    = {dblp computer science bibliography, https://dblp.org}
}

@inproceedings{NOAS16,
author = "Erika Nazaruka and Viktoria Ovchinnikova and Gundars Alksnis and Uldis Sukovskis",
title = "Veriﬁcation of BPMN Model Functional Completeness by using the Topological Functioning Model",
booktitle = "11th International Conference on Evaluation of Novel Software Approaches to Software Engineering (ENASE 2016)",
pages = "349--358",
year  = "2016",
ISBN = "978-989-758-189-2"
}

@inproceedings{BS20,
author = {Boonmepipit, Boodsarin and Suwannasart, Taratip},
title = {Test Case Generation from BPMN with DMN},
year = {2020},
isbn = {9781450376495},
publisher = {Association for Computing Machinery},
address = {New York, NY, USA},
ignoreurl = {https://doi.org/10.1145/3374549.3374582},
doi = {10.1145/3374549.3374582},
booktitle = {Proceedings of the 2019 3rd International Conference on Software and E-Business},
pages = {92–96},
numpages = {5},
location = {Tokyo, Japan},
series = {ICSEB '19}
}

@INPROCEEDINGS{FPPR12,
  author={Falcioni, Damiano and Polini, Andrea and Polzonetti, Alberto and Re, Barbara},
  booktitle={2012 12th International Conference on Quality Software}, 
  title={Direct Verification of BPMN Processes through an Optimized Unfolding Technique}, 
  year={2012},
  volume={},
  number={},
  pages={179-188},
  doi={10.1109/QSIC.2012.59}
}

@inproceedings{ARK19,
title = "A SysML-based Approach to Requirements Traceability Using BPMN and DMN",
author = "Corina Abdelahad and Daniel Riesco and Carlos Kavka",
booktitle = "ICSEA 2019, The Fourteenth International Conference on Software Engineering Advances",
pages = "210--216",
year = "2019",
ISBN = "978-1-61208-752-8"
}

@INPROCEEDINGS{Tak08,
  author={Takemura, Tsukasa},
  booktitle={2008 IEEE Asia-Pacific Services Computing Conference}, 
  title={Formal Semantics and Verification of BPMN Transaction and Compensation}, 
  year={2008},
  volume={},
  number={},
  pages={284-290},
  doi={10.1109/APSCC.2008.208}
  }

@INPROCEEDINGS{BMPP21,
  author={Bianchi, AnnaMaria and Mortari, Matteo and Pintavalle, Claudio and Pozzi, Giuseppe},
  booktitle={2021 IEEE International Conference on Digital Health (ICDH)}, 
  title={Putting BPMN and DMN to Work: a Pediatric Surgery Case Study}, 
  year={2021},
  volume={},
  number={},
  pages={154-159},
  doi={10.1109/ICDH52753.2021.00028}
  }

@article{CFPRTV21,
author = "Flavio Corradini and Fabrizio Fornari and Andrea Polini and Barbara Re and Francesco Tiezzi and Andrea Vandin",
title = "A formal approach for the analysis of BPMN collaboration models",
journal = "Journal of Systems and Software",
Volume = "180",
year = "2021",
doi = "https://doi.org/10.1016/j.jss.2021.111007"
}

@InProceedings{GCD21,
author="Groh{\'e}, Carl-Christian
and Corea, Carl
and Delfmann, Patrick",
editor="Polyvyanyy, Artem
and Wynn, Moe Thandar
and Van Looy, Amy
and Reichert, Manfred",
title="DMN 1.0 Verification Capabilities: An Analysis of Current Tool Support",
booktitle="Business Process Management Forum",
year="2021",
publisher="Springer International Publishing",
address="Cham",
pages="37--53",
isbn="978-3-030-85440-9"
}

@ARTICLE{DVT19,
  author={Dechsupa, C. and Vatanawood, W. and Thongtak, A.},
  journal={IEEE Access}, 
  title={Hierarchical Verification for the BPMN Design Model Using State Space Analysis}, 
  year={2019},
  volume={7},
  number={},
  pages={16795-16815},
  doi={10.1109/ACCESS.2019.2892958}
  }

@INPROCEEDINGS{SSM15,
  author={Solaiman, Ellis and Sun, Wenzhong and Molina-Jimenez, Carlos},
  booktitle={2015 IEEE International Conference on Services Computing}, 
  title={A Tool for the Automatic Verification of BPMN Choreographies}, 
  year={2015},
  volume={},
  number={},
  pages={728-735},
  doi={10.1109/SCC.2015.103}
  }

@inproceedings{HSS20,
author = {Hasi\'{c}, Faruk and Serral, Estefan\'{\i}a and Snoeck, Monique},
title = {Comparing BPMN to BPMN + DMN for IoT process modelling: a case-based inquiry},
year = {2020},
isbn = {9781450368667},
publisher = {Association for Computing Machinery},
address = {New York, NY, USA},
ignoreurl = {https://doi.org/10.1145/3341105.3373881},
doi = {10.1145/3341105.3373881},
booktitle = {Proceedings of the 35th Annual ACM Symposium on Applied Computing},
pages = {53–60},
numpages = {8},
location = {Brno, Czech Republic},
series = {SAC '20}
}

@article{HEB21,
title = {Code compliance checking of railway designs by integrating BIM, BPMN and DMN},
journal = {Automation in Construction},
volume = {121},
pages = {103427},
year = {2021},
issn = {0926-5805},
doi = {https://doi.org/10.1016/j.autcon.2020.103427},
ignoreurl = {https://www.sciencedirect.com/science/article/pii/S0926580520310074},
author = {Marco Haussler and Sebastian Esser and Andre Borrmann},
}

@article{HBPQK22,
title = {A First-Order Logic verification framework for communication-parametric and time-aware BPMN collaborations},
journal = {Information Systems},
volume = {104},
pages = {101765},
year = {2022},
issn = {0306-4379},
doi = {https://doi.org/10.1016/j.is.2021.101765},
ignoreurl = {https://www.sciencedirect.com/science/article/pii/S0306437921000272},
author = {Sara Houhou and Souheib Baarir and Pascal Poizat and Philippe Queinnec and Laid Kahloul},
}

@article{GLMS13,
  author       = {Hubert Garavel and
                  Fr{\'{e}}d{\'{e}}ric Lang and
                  Radu Mateescu and
                  Wendelin Serwe},
  title        = {{CADP} 2011: a toolbox for the construction and analysis of distributed
                  processes},
  journal      = {Int. J. Softw. Tools Technol. Transf.},
  volume       = {15},
  number       = {2},
  pages        = {89--107},
  year         = {2013},
  url          = {https://doi.org/10.1007/s10009-012-0244-z},
  doi          = {10.1007/S10009-012-0244-Z},
  bibsource    = {dblp computer science bibliography, https://dblp.org}
}

@InProceedings{BEM13,
  author =	{Bae, Kyungmin and Escobar, Santiago and Meseguer, Jos\'{e}},
  title =	{{Abstract Logical Model Checking of Infinite-State Systems Using Narrowing}},
  booktitle =	{24th International Conference on Rewriting Techniques and Applications (RTA 2013)},
  pages =	{81--96},
  series =	{Leibniz International Proceedings in Informatics (LIPIcs)},
  ISBN =	{978-3-939897-53-8},
  ISSN =	{1868-8969},
  year =	{2013},
  volume =	{21},
  editor =	{van Raamsdonk, Femke},
  publisher =	{Schloss Dagstuhl -- Leibniz-Zentrum f{\"u}r Informatik},
  address =	{Dagstuhl, Germany},
  URL =		{https://drops.dagstuhl.de/entities/document/10.4230/LIPIcs.RTA.2013.81},
  URN =		{urn:nbn:de:0030-drops-40554},
  doi =		{10.4230/LIPIcs.RTA.2013.81}
}

@InProceedings{VGLC22,
author="Vandin, Andrea
and Giachini, Daniele
and Lamperti, Francesco
and Chiaromonte, Francesca",
editor="Bowles, Juliana
and Broccia, Giovanna
and Pellungrini, Roberto",
title="MultiVeStA: Statistical Analysis of Economic Agent-Based Models by Statistical Model Checking",
booktitle="From Data to Models and Back",
year="2022",
publisher="Springer International Publishing",
address="Cham",
pages="3--6",
isbn="978-3-031-16011-0"
}

@INPROCEEDINGS{camundamodeler,
  author={David, Ghiurau and Zmaranda, Doina Rodica and Gyorodi, Robert-Stefan and Gyorodi, Cornelia Aurora},
  booktitle={2023 17th International Conference on Engineering of Modern Electric Systems (EMES)}, 
  title={Exploring the Impact of Workflow Engines on Business Process Management in Enterprise Applications. A case-study: Camunda}, 
  year={2023},
  volume={},
  number={},
  pages={1-4},
  doi={10.1109/EMES58375.2023.10171706}}

@book{bpmn_book,
  title={The Process: Business Process Modeling Using BPMN},
  author={Grosskopf, A. and Decker, G. and Weske, M.},
  isbn={9780929652269},
  lccn={2009923535},
  year={2009},
  publisher={Meghan-Kiffer Press}
}

@article{CCFR23,
author = {Compagnucci, Ivan and Corradini, Flavio and Fornari, Fabrizio and Re, Barbara},
year = {2023},
month = {06},
pages = {1-24},
title = {A Study on the Usage of the BPMN Notation for Designing Process Collaboration, Choreography, and Conversation Models},
volume = {66},
journal = {Business \& Information Systems Engineering},
doi = {10.1007/s12599-023-00818-7}
}

@INPROCEEDINGS{DFM16,
  author={Dangarska, Zhivka and Figl, Kathrin and Mendling, Jan},
  booktitle={2016 IEEE 20th International Enterprise Distributed Object Computing Workshop (EDOCW)}, 
  title={An Explorative Analysis of the Notational Characteristics of the Decision Model and Notation (DMN)}, 
  year={2016},
  volume={},
  number={},
  pages={1-9},
  doi={10.1109/EDOCW.2016.7584345}
  }

@ARTICLE{RW85,
  author={Rapps, S. and Weyuker, E.J.},
  journal={IEEE Transactions on Software Engineering}, 
  title={Selecting Software Test Data Using Data Flow Information}, 
  year={1985},
  volume={SE-11},
  number={4},
  pages={367-375},
  doi={10.1109/TSE.1985.232226}
  }

@inproceedings{WJD13,
author = {Wang, Yi and Jiang, Tong Hai and Dong, Jun},
title = {Fast Loop Boundary Coverage Testing in Symbolic Execution},
year = {2013},
month = {8},
volume = {718},
pages = {2314--2317},
booktitle = {Advanced Measurement and Test III},
series = {Advanced Materials Research},
publisher = {Trans Tech Publications Ltd},
doi = {10.4028/www.scientific.net/AMR.718-720.2314}
}

@book{Hay01,
  title={A Practical Tutorial on Modified Condition/Decision Coverage},
  author={Hayhurst, K.J.},
  isbn={9781428995994},
  url={https://books.google.it/books?id=aqMz3xtU6HsC},
  year={2001},
  publisher={DIANE Publishing}
}

@book{San2011,
place={Cambridge},
title={Introduction to Bisimulation and Coinduction},
publisher={Cambridge University Press},
author={Sangiorgi, Davide},
year={2011}
}

@article{OFRMM12,
title = {Making sense of business process descriptions: An experimental comparison of graphical and textual notations},
journal = {Journal of Systems and Software},
volume = {85},
number = {3},
pages = {596-606},
year = {2012},
note = {Novel approaches in the design and implementation of systems/software architecture},
issn = {0164-1212},
doi = {https://doi.org/10.1016/j.jss.2011.09.023},
url = {https://www.sciencedirect.com/science/article/pii/S0164121211002408},
author = {Avner Ottensooser and Alan Fekete and Hajo A. Reijers and Jan Mendling and Con Menictas}
}

@article{GHLW18,
title = {BPMN 2.0: The state of support and implementation},
journal = {Future Generation Computer Systems},
volume = {80},
pages = {250-262},
year = {2018},
issn = {0167-739X},
doi = {https://doi.org/10.1016/j.future.2017.01.006},
url = {https://www.sciencedirect.com/science/article/pii/S0167739X17300250},
author = {Matthias Geiger and Simon Harrer and Jörg Lenhard and Guido Wirtz}
}

@article{FDAM12,
title = {Automatic execution of business process models: Exploiting the benefits of Model-driven Engineering approaches},
journal = {Journal of Systems and Software},
volume = {85},
number = {3},
pages = {607-625},
year = {2012},
note = {Novel approaches in the design and implementation of systems/software architecture},
issn = {0164-1212},
doi = {https://doi.org/10.1016/j.jss.2011.09.022},
url = {https://www.sciencedirect.com/science/article/pii/S0164121211002391},
author = {J. Fabra and V. {De Castro} and P. Álvarez and E. Marcos}
}

@article{PFP19,
title = {Business process model refactoring applying IBUPROFEN. An industrial evaluation},
journal = {Journal of Systems and Software},
volume = {147},
pages = {86-103},
year = {2019},
issn = {0164-1212},
doi = {https://doi.org/10.1016/j.jss.2018.10.012},
url = {https://www.sciencedirect.com/science/article/pii/S016412121830222X},
author = {Ricardo Pérez-Castillo and María Fernández-Ropero and Mario Piattini}
}

@article{CMRRT22,
title = {Formalising and animating multiple instances in BPMN collaborations},
journal = {Information Systems},
volume = {103},
pages = {101459},
year = {2022},
issn = {0306-4379},
doi = {https://doi.org/10.1016/j.is.2019.101459},
url = {https://www.sciencedirect.com/science/article/pii/S0306437919305113},
author = {Flavio Corradini and Chiara Muzi and Barbara Re and Lorenzo Rossi and Francesco Tiezzi}
}

@article{VTS22,
title = {Modelling and executing IoT-enhanced business processes through BPMN and microservices},
journal = {Journal of Systems and Software},
volume = {184},
pages = {111139},
year = {2022},
issn = {0164-1212},
doi = {https://doi.org/10.1016/j.jss.2021.111139},
url = {https://www.sciencedirect.com/science/article/pii/S0164121221002363},
author = {Pedro Valderas and Victoria Torres and Estefanía Serral}
}

@article{NQT25,
title = {Runtime defect prediction of industrial business processes: A focused look at real-life SAP systems},
journal = {Journal of Systems and Software},
volume = {222},
pages = {112306},
year = {2025},
issn = {0164-1212},
doi = {https://doi.org/10.1016/j.jss.2024.112306},
url = {https://www.sciencedirect.com/science/article/pii/S0164121224003509},
author = {Max Nijholt and Giovanni Quattrocchi and Damian Andrew Tamburri}
}

@InProceedings{CKD24,
author="Corea, Carl
and Kampik, Timotheus
and Delfmann, Patrick",
editor="De Weerdt, Jochen
and Pufahl, Luise",
title="Empirical Evidence of DMN Errors in the Wild - An SAP Signavio Case Study",
booktitle="Business Process Management Workshops",
year="2024",
publisher="Springer Nature Switzerland",
address="Cham",
pages="326--336",
isbn="978-3-031-50974-2"
}

@InProceedings{KGR22,
author="Kirikkayis, Yusuf
and Gallik, Florian
and Reichert, Manfred",
editor="Almeida, Jo{\~a}o Paulo A.
and Karastoyanova, Dimka
and Guizzardi, Giancarlo
and Montali, Marco
and Maggi, Fabrizio Maria
and Fonseca, Claudenir M.",
title="Modeling, Executing and Monitoring IoT-Driven Business Rules with BPMN and DMN: Current Support and Challenges",
booktitle="Enterprise Design, Operations, and Computing",
year="2022",
publisher="Springer International Publishing",
address="Cham",
pages="111--127",
isbn="978-3-031-17604-3"
}

@book{mc_book,
author = {Clarke, Edmund M. and Henzinger, Thomas A. and Veith, Helmut and Bloem, Roderick},
title = {Handbook of Model Checking},
year = {2018},
isbn = {3319105744},
publisher = {Springer Publishing Company, Incorporated},
edition = {1st}
}
